\documentclass[12pt]{article}
\usepackage{float}
\usepackage{physics}
\usepackage{subfig}
\usepackage[nottoc]{tocbibind}
\usepackage{authblk}
\usepackage{amsmath}
\usepackage{cite}
\usepackage{cite}
\usepackage{lineno}
\usepackage{xcolor}
\usepackage{lineno}
\usepackage{ulem}
\usepackage{amssymb}
\usepackage{graphicx}
\usepackage{slashed}
\usepackage{comment}
\usepackage{epsfig}
\usepackage{tikzsymbols}
\usepackage{appendix}
\usepackage{epstopdf}
\usepackage{upgreek}
\usepackage[numbers,sort&compress]{natbib}

\usepackage{xcolor}
\usepackage[compat=1.1.0]{tikz-feynman}
\usepackage{tikzsymbols}
\usepackage[colorlinks=true,urlcolor=black,linkcolor=black, citecolor=black]{hyperref}
\usepackage{hyperref}
\usepackage{xcolor}

\hypersetup{
    colorlinks=true,
    linkcolor=red,   
    citecolor=blue,  
    urlcolor=magenta
}

\usepackage[a4paper, total={6.5in, 10.5in}]{geometry}
\usepackage{mathtools,slashed}
\usepackage{tikz}
\definecolor{deepmagenta}{rgb}{0.8, 0.0, 0.8}
\definecolor{mediumtealblue}{rgb}{0.0, 0.33, 0.71}
\definecolor{warmblack}{rgb}{0.0, 0.26, 0.26}
\definecolor{bostonuniversityred}{rgb}{0.8, 0.0, 0.0}
\definecolor{junglegreen}{rgb}{0.16, 0.67, 0.53}
\definecolor{lightcornflowerblue}{rgb}{0.6, 0.81, 0.93}

\title{\textbf{ Towards Testable Type-III Leptogenesis in Non-Standard Early Universe Scenarios } }

\author[1]{\small{Simran Arora \thanks{009simranarora@gmail.com}}}

\affil[1]{\it{\small{Department of Physics and Astronomical Science, Central University of Himachal Pradesh, Dharamshala 176215,
INDIA.}}}

\author[2]{Devabrat Mahanta \thanks{devabrat@pragjyotishcollege.ac.in}}
\affil[2]{\it{Department of Physics, Pragjyotish College, Guwahati 781009, INDIA }}

\begin{document}

\maketitle
\begin{abstract}
Leptogenesis is an elegant way to explain the baryon asymmetry of the Universe in connection to the neutrino mass and mixing. Although leptogenesis from the decay of a heavy Majorana neutrino has been the minimal set up, it is also motivating to look for leptogenesis from the decay of triplet fermion as it can have detectable signatures in the experiments. However, due to strong gauge annihilations and constraints from neutrino sector, the triplet fermions have to be as heavy as $10^{10}$ GeV or more to generate the observed baryon asymmetry. While this prediction is based on the standard radiation dominated history of the early Universe, it is also possible to have a non-standard expansion history of the Universe prior to the big-bang nucleosynthesis. In this work we study triplet leptogenesis in two non-standard cosmological scenarios, where the Universe expands faster than radiation and a scalar tensor theory of gravity. We show that it is possible to have successful leptogenesis with a few TeV triplet fermion for fast expanding Universe and a few hundered TeV for a scalar tensor gravity theory. 

\end{abstract}

\newpage

\tableofcontents

\newpage

\section{Introduction}
\label{section1}

It is observed that our Universe is dominated by matter over antimatter. This asymmetry between matter and antimatter is expressed in terms of the Baryon to Photon ratio defined by $\eta_{B}=\left( n_{B}-n_{\bar{B}} \right)/n_{\gamma}$. The recent measurement of $\eta_{B}$ by the Planck satellite report a value  $\eta_{B}=(6.21\pm 0.16)\times 10^{-10}$ \cite{Planck:2018vyg}. While the measurement is also consistent with the estimates from big bang nucleosynthesis (BBN), the origin of the asymmetry has been a long standing problem in particle physics and cosmology. To dynamically generate a baryon asymmetry in the Universe three conditions known as the Sakharov conditions \cite{Sakharov:1967dj} must be satisfied. These conditions are (i) Baryon number violation (ii) C and CP violation and (iii) Departure from thermal equilibrium. Although the standard model of particle physics with an expanding Universe has all the necessary ingredients to satisfy the Sakharov conditions, the $\rm CP$ violation present in the SM in not adequate to generate the observed asymmetry. Among the different beyond the standard model (BSM) scenarios,  leptogenesis is an interesting way of generating the observed asymmetry in connection with neutrino masses \cite{Fukugita:1986hr,Minkowski:1977sc, Mohapatra:1980yp}. In conventional leptogenesis scenarios a net lepton asymmetry is generated from the decay of singlet Majorana right handed neutrino. The lepton asymmetry later gets converted into a baryon asymmetry by the \textit{sphaleron} process \cite{Kuzmin:1985mm}. One such simple extension of the SM is to add two or three copies of Majorana right handed neutrinos ($N_{i}$) \cite{Minkowski:1977sc,Yanagida:1980xy}. They not only generate tiny masses for the light neutrinos but produces a $\rm B-L$ asymmetry from the decay of the lightest right handed neutrino (RHN) $N_{1}\longrightarrow l H$. The required CP violation is generated from the interference of tree level and one loop diagram for the decay $N_{i}\longrightarrow l H$.

In the type-I seesaw model, there exist a lower bound on the mass of the lightest right handed neutrino, $M_{1} \gtrsim 10^{9}$ GeV known as the Davidson-Ibarra (DI) bound \cite{Davidson_2002}. With RHN below this mass, sufficient $\rm B-L$ asymmetry can not be produced. Consequently, such heavy right-handed neutrinos remain far beyond the reach of current and foreseeable experimental probes. Besides, the RHNs in the type-I seesaw model are singlet under $SU(2)_{L}$, and therefore have no SM gauge interaction with the SM particles. Therefore, it is difficult to probe such particles at collider experiments. Another simple extension of the SM to explain the tiny neutrino masses is to add two or three copies of vector like triplet fermions ($\Sigma_{i}$) \cite{Foot:1988aq}. A net lepton asymmetry can be generated from the out-of equilibrium decay of the neutral component of the lightest triplet fermion ($\Sigma_{1}$) \cite{Ma:1998dx, Hambye:2012fh, Vatsyayan:2022rth, Arora:2024lpq}. In this scenario due to the strong gauge annihilations, it is difficult to achieve the out of equilibrium condition for the triplet. As a result the mass required for the lightest triplet fermion $M_{\Sigma_1}\gtrsim 10^{10}$ GeV, is even higher than the type-I model. In \cite{Vatsyayan:2022rth}, the authors shown that the lower limit on mass of triplet fermion can be lowered to $M_{\Sigma_{1}}\gtrsim 10^{7}$ GeV by adding an additional Higgs doublet with hypercharge one \cite{Vatsyayan:2022rth}. However, a triplet fermion as heavy as $M_{\Sigma_1} \gtrsim 10^{7}$ GeV is still far beyond the detectable range of near future experiments. 

 Such lower bound on the scale of leptogenesis is applicable for standard radiation-dominated Universe. There exist no experimental evidence to suggest that the Universe is radiation dominated prior to the BBN which is nearly 1 s after the big bang. The prediction of leptogenesis depends on the background Universe and it's evolution. There have been several works on leptogenesis in non-standard cosmology in the context of vanilla leptogenesis \cite{Mahanta:2019sfo, Chen:2019etb, DiMarco:2022doy,JyotiDas:2021shi, Chang:2021ose,Chakraborty:2022gob}. In \cite{Mahanta:2019sfo}, the authors have studied leptogenesis in a fast expanding Universe (FEU) and a early matter dominated (EMD) Universe in the context of scotogenic model \cite{Ma:2006km}. In the scotogenic model due to the radiative mass generation for the neutrinos the Yukawa coupling for the lightest RHN can be small making washout processes in leptogenesis very weak. It is shown that in such a weak washout regime of vanilla leptogenesis a fast expanding Universe make the asymmetry production slower requiring a higher value of the decaying right handed neutrino. 
 However, in the case of strong washout leptogenesis scenarios such as in the type-I seesaw model, the departure from thermal equilibrium can be significantly enhanced in the case of a fast expanding Universe. In \cite{Chen:2019etb,Chakraborty:2022gob} it is shown that it results in the possibility to significantly lowering the scale of leptogenesis.  
 
 In \cite{Mahanta:2022gsi}, the authors have discussed the consequences of three different non-standard cosmological backgrounds, (i) a FEU, (ii) an EMD Universe and (iii) a scalar tensor theory of gravity (STG) for a WIMPy leptogenesis scenario. In WIMPy leptogenesis, a net $\rm B-L$ asymmetry is generated from the scattering of weakly interacting massive particle (WIMP) dark matter (DM). In a FEU, the WIMP particles goes out-of equilibrium early, overproducing their abundance. Due to this enhancement in out-of-equilibrium abundance for the WIMP the $\rm B-L$ asymmetry production from their annihilation also increases. In an EMD Universe, the entropy injection from the matter field decay always dilutes the asymmetry produced requiring a higher scale of leptogenesis compared to the standard radiation dominated Universe.

 Motivating from this, we study the effects of three different non-standard cosmologies. where the Universe expands faster than radiation on triplet leptogenesis in this work. On the other hand in \cite{Mahanta:2019sfo,Mahanta:2022gsi}, it is shown that in an EMD Universe due to the entropy dilution from the matter field the scale of leptogenesis is always significantly higher compared to the standard radiation dominated Universe. Since it is not phenomenologically motivating we do not study this scenario here. The paper is organized as follows: In section \ref{section2}, we discuss the minimal type-III seesaw model and analyze the triplet leptogenesis in standard cosmology, in section \ref{section3} we show the details of triplet leptogenesis in a FEU and in section \ref{section4}, we consider triplet leptogenesis in a modified theory of gravity. In section \ref{section5} we conclude summarizing the results.

\section{Leptogenesis in type-III seesaw}
\label{section2}

We consider the minimal type-III seesaw model with two copies of vector like $SU(2)_{L}$ triplet fermion ($\Sigma_{i}$) \cite{Foot:1988aq,Ma:1998dn}. The relevant Lagrangian for the triplets can be written as 


\begin{eqnarray}
    \mathcal{L} \subset \sum_{i=1}^{2} \rm Tr [\Sigma_{i} \slashed{D}\Sigma_{i}]- \frac{1}{2} \left[ \bar{\Sigma}_{i}(M_{\Sigma})_{ij}\Sigma^{c}_{j} + \bar{\Sigma^{c}_{i}}(M_{\Sigma}^{*})_{ij}\Sigma_{j}\right]-\sqrt{2}(Y_{\Sigma})_{i\alpha}\tilde{H^{\dagger}}\bar{\Sigma_{i}} L_{\alpha}-\sqrt{2}(Y_{\Sigma}^{\dagger})_{\alpha i}\bar{L}_{\alpha}\Sigma_{i} \tilde{H},
\end{eqnarray}

\noindent where $(M_{\Sigma})_{ij}$ are the elements of the mass matrix of $\Sigma_{i}$ and $(Y_{\Sigma})_{i\alpha}$ are the elements of the Yukawa matrix $Y_{\Sigma}$. Here, $L_{\alpha}=(\nu_{\alpha}, l_{\alpha})^{T}$, with $\alpha=e,\mu,\tau$ is the SM lepton doublet. $H=(H^{+},H^{0})^{T}$ is the SM Higgs doublet and $\tilde{H}=i \sigma_{2}H^{*}$. $D_{\mu}=\partial_{\mu}-ig\tau^{A}W_{\mu}^{A}/2$ is the co-variant derivative for the triplet fermions $\Sigma_{i}$. In the adjoint representation the triplet fermion is written as 

\begin{eqnarray}
    \Sigma= \begin{pmatrix}
        \Sigma^{0}/\sqrt{2} && \Sigma^{+} \\
        \Sigma^{-}  && -\Sigma^{0}/\sqrt{2}
    \end{pmatrix}.
\end{eqnarray}

\noindent After the electro-weak symmetry breaking, the neutrinos get mass and the corresponding mass matrix can be written as

\begin{eqnarray}
    M_{\nu}= -\dfrac{v^{2}}{2} Y_{\Sigma}^{T} M_{\Sigma}^{-1}Y_{\Sigma},
\end{eqnarray}

\noindent where $v$ is the vacuum expectation value (VEV) of the SM Higgs, $Y_{\Sigma}$ is the triplet fermion Yukawa matrix and $M_{\Sigma}$ is the triplet fermion mass matrix in the diagonal basis. The Casas-Ibarra (CI) \cite{Casas:2001sr} parameterization parametrizes the Yukawa matrix consistent $Y_{\Sigma}$, in terms of the light neutrino masses and mixing parameters. Implementing the CI parameters we write the Yukawa matrix $Y_{\Sigma}$ as 

\begin{equation}
   Y_{\Sigma} = U D_{\sqrt{M_{\nu}}} R^{\dagger} D_{\sqrt{M_{\Sigma}}}.   
\end{equation}

Here, $U$ is the Pontecorvo–Maki–Nakagawa–Sakata (PMNS) neutrino mixing matrix. $D_{\sqrt{M_{\nu}}}=diag(\sqrt{m_{1}},\sqrt{m_{2}},\sqrt{m_{3}})$ is a diagonal matrix with $m_{i}$s being the light neutrino masses. $R$ is a complex orthogonal matrix and $D_{\sqrt{M_{\Sigma}}}=diag(\sqrt{M_{\Sigma_{1}}},\sqrt{M_{\Sigma_{2}}})$ is a diagonal matrix where $M_{\Sigma_{i}}$ are the masses of the triplet fermions.

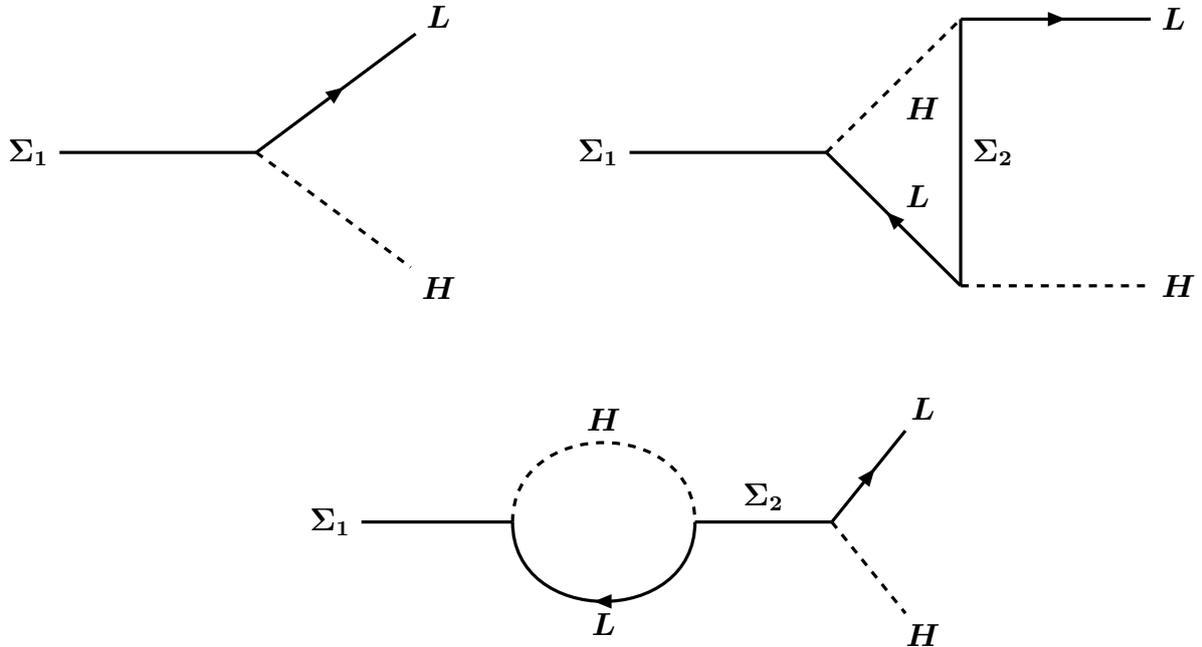
\begin{figure}
\begin{tikzpicture}
[scale=0.6]
    \begin{feynman}
        \vertex (i1) at (-5, 0) {\(\boldsymbol{\Sigma_{1}}\)};
        \vertex (v1) at (0, 0);
        \vertex (o1) at (4, 3) {\(\boldsymbol{L}\)};
        \vertex (o2) at (4, -3) {\(\boldsymbol{H}\)};
        
        \diagram* {
          (i1) -- [plain,very thick] (v1),
          (v1) -- [fermion, very thick] (o1),
          (v1) -- [scalar,very thick] (o2),
        };
      \end{feynman}
  \hspace{4.5cm}
\begin{feynman}
[scale=0.5]
    \vertex (i) {\(\boldsymbol{\Sigma_1}\)};
    \vertex [right=3cm of i] (v1);

    \vertex [above right=2.5cm of v1] (v2) ;
    \vertex [below right=2.5cm of v1] (v3);

    \vertex [right=2.5cm of v2] (f1) {\(\boldsymbol{L}\)};
    \vertex [right=2.5cm of v3] (f2) {\(\boldsymbol{H}\)};

    \diagram* {
        (i) -- [plain, very thick] (v1),
        
        (v1) -- [scalar, very thick, edge label'=\(\boldsymbol{H}\)] (v2) -- 
        [plain, very thick, edge label=\(\boldsymbol{\Sigma_2}\)] (v3) -- 
        [fermion,very thick, edge label'=\(\boldsymbol{L}\)] (v1),

        (v2) -- [fermion, very thick] (f1),
        (v3) -- [scalar, very thick] (f2),
    };
\end{feynman}
\end{tikzpicture}

\vspace{1cm}
\centering
      \begin{tikzpicture}[scale=0.6]
      \begin{feynman}
        \vertex (i1) at (-4, 0) {\(\boldsymbol{\Sigma_{1}}\)};
        \vertex (v1) at (0, 0);
        \vertex (loop1) at (4, 0); 
        \vertex (loop2) at (4, 5); 
        \vertex (v2) at (4, 0);
        \vertex (v3) at (7,0);
        \vertex (o1) at (9, 2.5) {\(\boldsymbol{L}\)};
        \vertex (o2) at (9, -2.5) {\(\boldsymbol{H}\)};
        
        \diagram* {
          (i1) -- [plain,very thick] (v1) -- [white,plain] (v2)  -- [plain,very thick,edge label=\(\boldsymbol{{\Sigma_{2}}}\)] (v3)-- [fermion,very thick] (o1),
          (v3) -- [scalar,very thick] (o2),
          (v1) -- [scalar,very thick, half left, edge label=\(\boldsymbol{{H}}\)] (loop1) -- [fermion, very thick,half left, edge label=\(\boldsymbol{L}\)] (v1) 
        };
      \end{feynman}
    \end{tikzpicture}

    \caption{Diagrams relevant for CP asymmetry by the decay of lightest triplet fermion.}
    \label{fig:feynmann}
\end{figure}

In type III seesaw model a net lepton asymmetry can be generated from the out-of-equilibrium decay of the lightest triplet fermion $\Sigma_1$.  The required CP asymmetry can be generated from the interference of tree level, one loop vertex diagram and self-energy diagram shown in Fig. (\ref{fig:feynmann}). The CP asymmetry  generated by the lightest triplet fermion is defined by the CP asymmetry parameter $\epsilon_1$ and is given by

\begin{eqnarray}
\epsilon_{1}
=-
\frac{3 M_{\Sigma_1}}{2 M_{\Sigma_2}}
\frac{\Gamma_2}{M_{\Sigma_2}}
I_2
\frac{V_2 + 2 S_2}{3} ,
\end{eqnarray}

\noindent where

\begin{eqnarray}
    I_2 =
\frac{
\Im\!\left[(Y_{\Sigma}^\dagger Y_{\Sigma})^{2}_{12}\right]
}{
|(Y_{\Sigma}^\dagger Y_{\Sigma})_{11}|\; |(Y_{\Sigma}^\dagger Y_{\Sigma})_{22}|
}.
\end{eqnarray}

\noindent The $S_2$ and $V_2$ are the loop factors coming from the self energy and vertex correction diagrams, respectively, and are given by 

\begin{eqnarray}
    S_{2} & = & \frac{M_{\Sigma_{2}}^{2} \Delta M_{21}^{2}}{(\Delta M_{21}^{2})^{2}+M_{\Sigma_{1}}^{2}\Gamma_{2}^{2}}, \\
    V_{2} & = & 2 \frac{M_{\Sigma_{2}}^{2}}{M_{\Sigma_{1}}^{2}} \left[  \left(1+\frac{M_{\Sigma_{2}}^{2}}{M_{\Sigma_{1}}^{2}} \right)\text{log} \left( 1+\frac{M_{\Sigma_{1}}^{2}}{M_{\Sigma_{2}}^{2}} \right) -1  \right].
\end{eqnarray}

The same Yukawa coupling that generates neutrino mass also generates the $\rm B-L$ asymmetry. There exists an upper bound known as the Davidson-Ibarra (DI) bound on the CP asymmetry produced \cite{Davidson:2002qv,Hambye:2003rt}. Similar to the type-I seesaw the DI bound for the type-III seesaw model can be derived to be \cite{Davidson:2002qv,Baldes:2025uwz}

\begin{eqnarray}
    \epsilon_{1}\lesssim \dfrac{3}{16\pi}  \dfrac{M_{\Sigma_{1}}}{v^{2}}(m_{3}-m_{1}). 
\end{eqnarray}

For a standard radiation dominated Universe the CP asymmetry parameter $\epsilon_{1}\gtrsim 10^{-6}$ to generate the observed baryon asymmetry. Therefore, to generate the observed baryon asymmetry, the lightest triplet mass has to be $M_{\Sigma_{1}}\gtrsim 3\times 10^{10} \rm GeV$ \cite{Hambye:2003rt}. The presence of such heavy particles can lead to large quadratic correction to the Higgs mass. Besides, such a heavy triplet fermion is far beyond the experimental range of current and future experiments. Therefore, it is motivating to look for leptogenesis scenario where adequate asymmetry can be generated from a light $M_{\Sigma_{1}}\sim \mathcal{O}(1$ TeV), triplet fermion. The current searches of triplet fermion at the ATLAS \cite{ATLAS:2020wop} detector of the LHC experiment has put a lower limit on the triplet mass to be $790$ GeV at $95\%$ C.L. Similarly, the lower limit from  the CMS \cite{CMS:2019lwf} detectors at $13$ TeV LHC is $880$ GeV at $95\%$ C.L. There are attempts to reduce the lower bound on the mass of the lightest fermion triplet by adding extra scalars \cite{Clarke:2015hta,Alanne:2018brf,Hugle:2018qbw,Goswami:2021eqy,Bhattacharya:2024ohh,Borah:2025hpo}. In \cite{Vatsyayan:2022rth}, the authors have proposed a type III seesaw model where the presence of an additional Higgs doublet lowers the scale of leptogenesis to $M_{\Sigma_{1}}\gtrsim 10^{7} \rm GeV$. On the other hand a TeV scale leptogenesis can be achieved by resonantly enhancing the CP asymmetry from the self energy contribution \cite{Pilaftsis:2003gt,Arora:2024lpq}. Such a resonant enhancement is satisfied when $\mid M_{\Sigma_{j}}-M_{\Sigma_{i}} \mid \sim \Gamma_{i,j}/2$. In this work we do not restrict to such a fine tuned condition.

The DI bound is applicable when the background Universe has a standard radiation dominated history. However, prior to the big-bang nucleosynthesis (BBN) the Universe does not necessarily have to be dominated by standard radiation. In this work we study the possibility of lowering the scale of triplet leptogenesis in non-standard cosmological histories of the early Universe \cite{Fierz:1956zz, Brans:1961sx, DEramo:2017gpl}. In particular we study the effects of non-standard cosmologies on the gauge scatterings of the triplet fermion and the washout processes.

At first, we write Boltzmann equations for a triplet fermion $\Sigma_1$ considering its decay to a lepton and Higgs and its gauge annihilations in standard radiation dominated Universe

\begin{align}\label{Boltz1}
\nonumber
\frac{dY_{\Sigma_1}}{dz} &= -D_{\Sigma_1} (Y_{\Sigma_1}-Y^{eq}_{\Sigma_1})-S_{A}(Y_{\Sigma_1}^{2}-(Y_{\Sigma_1}^{eq})^{2}), \\
  \frac{dY_{B-L}}{dz} &= -\epsilon D_{\Sigma_1} (Y_{\Sigma_1}-Y^{eq}_{\Sigma_1}) - W_{\Sigma_1} Y_{B-L} - W_{\Delta L} Y_{B-L}.  \\ \nonumber
\end{align}

\noindent The $Y_{\Sigma_{1}}$ and $Y_{B-L}$ are the co-moving number densities of $\Sigma_{1}$ and $\rm B-L$ asymmetry respectively and $z=M_{\Sigma_{1}}/T $ is a dimensionless inverse temperature scale. Here the decay term $D_{\Sigma_1}$ is given by 
\begin{align}
  D_{\Sigma_1} = K_{\Sigma_1}  \dfrac{\kappa_{1}(z)}{\kappa_{2}(z)}.
\end{align}
 Here, $\kappa_{i}$s are modified Bessel functions of second kind and $K_{\Sigma_1}$ is the decay parameter defined as  
 \begin{align}
    K_{\Sigma_1} =  \frac{\Gamma_{\Sigma_1}}{\textbf{H}(z=1)}.
 \end{align}
The decay width $\Gamma_{\Sigma_1}$ is given by 
\begin{eqnarray}
    \Gamma_{\Sigma_{1}} = \frac{M_{\Sigma_1}}{8 \pi}\left( Y_{\Sigma}^{\dagger}Y_{\Sigma}  \right).
\end{eqnarray}

\noindent and $\textbf{H}(z)$ is the Hubble parameter in the standard radiation dominated Universe and is given by 

\begin{equation}
    \textbf{H}(z)=\sqrt{\dfrac{8\pi^{3}g_{*}}{90}}\dfrac{M_{\Sigma_{1}}^{2}}{M_{Pl}^{2}}\dfrac{1}{z^{2}}.
\end{equation}

\noindent $g_{*}$ is the effective number of relativistic degrees of freedom present in the Universe and $M_{Pl}\simeq 1.22\times 10^{19}$ GeV is the Planck mass. The $W_{\Sigma_{1}}$ is the inverse decay term that can washout some of the generated asymmetry and is defined as 
\begin{eqnarray}
    W_{\Sigma_1}= \frac{1}{4}K_{\Sigma_{1}}z^{3}\kappa_{1}(z).
\end{eqnarray}
 The term $W_{\Delta L}$ takes care of the washouts coming from the lepton number violating scattering terms defined as $W_{\Delta L}=\Gamma_{scatterings}/\textbf{H}(z) z^{2}$. The important scatterings in this model are $L H\longrightarrow \overline{L} H^{*}$ and $LL\longrightarrow H H^{*}$.  We give the details of the scattering washout term in Appendix \ref{Appen:A}.
 $S_{A}$ consists of the gauge boson mediated annihilation rates for the fermion triplet $\Sigma_{1}$ and it is defined as

\begin{equation}
    S_{A} = \bigg(\frac{\pi^2 g^{*1/2}M_Pl}{1.66*180 g_{\Sigma_{1}}^2}\bigg)\frac{1}{M_{\Sigma_{1}}}\bigg(\frac{I_{z}}{z \kappa_{2}(z)^{2}}\bigg),
\end{equation}
where $g_{\Sigma_{1}}$ is the internal degrees of freedom of $\Sigma_{1}$, and $I_{z}$ is an integral consisting the thermal average of the annihilation cross section $\hat{\sigma}_{A}$ 

\begin{equation}
    I(z) =  \int_{4}^{\infty} \sqrt{x} \kappa (z \sqrt{x})\hat{\sigma}_{A}(x)dx .
\end{equation}
Here, the integration variable $x=s/M_{\Sigma_{1}^{2}}$, $s$ being centre of mass energy. The annihilation cross-section $\hat{\sigma}_{A}(x)$ is given by \cite{Biswas:2023azl} 
\begin{equation}
\hat{\sigma}_{A}(x) = \frac{6g^4}{72 \pi}\Bigg[\frac{45}{2}r(x)-\frac{27}{2}r(x)^3-\{9(r(x)^2-2)+18(r(x)^2-1)^2\}\text{ln}\bigg(\frac{1+r(x)}{1-r(x)}\bigg)\Bigg],   
\end{equation}
with $r(x) = \sqrt{1-4/x}$ and $g$ is the $SU(2)_{L}$ gauge coupling constant.   

\begin{figure}
    \centering
        \includegraphics[scale=0.45]{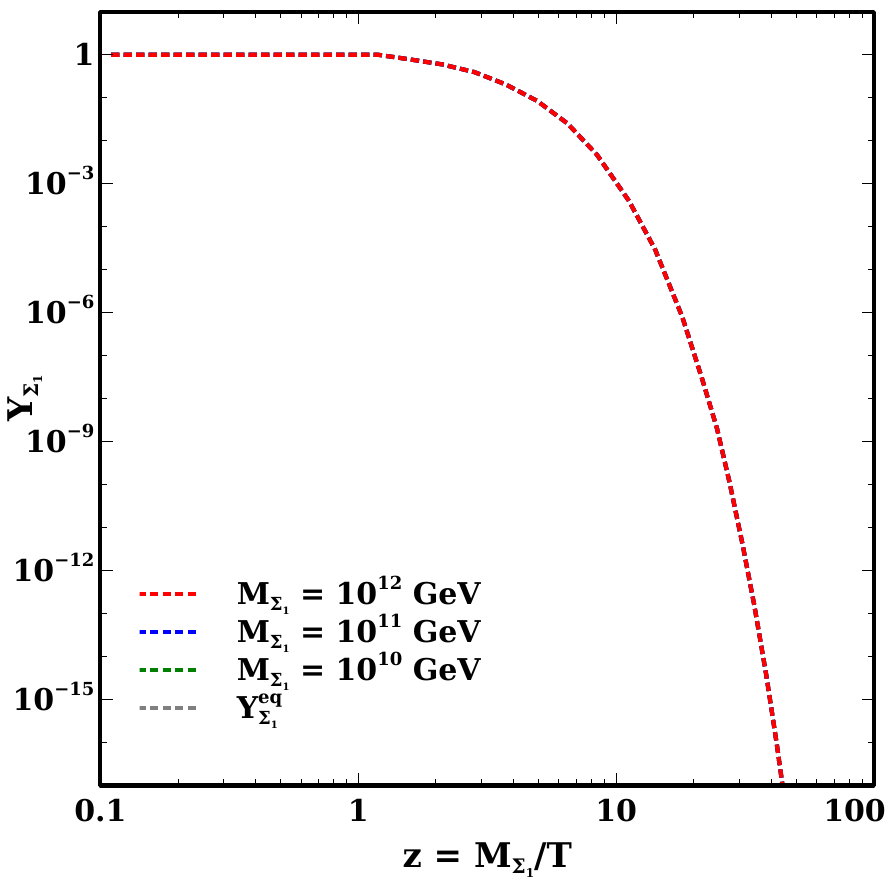}
    \includegraphics[scale=0.45]{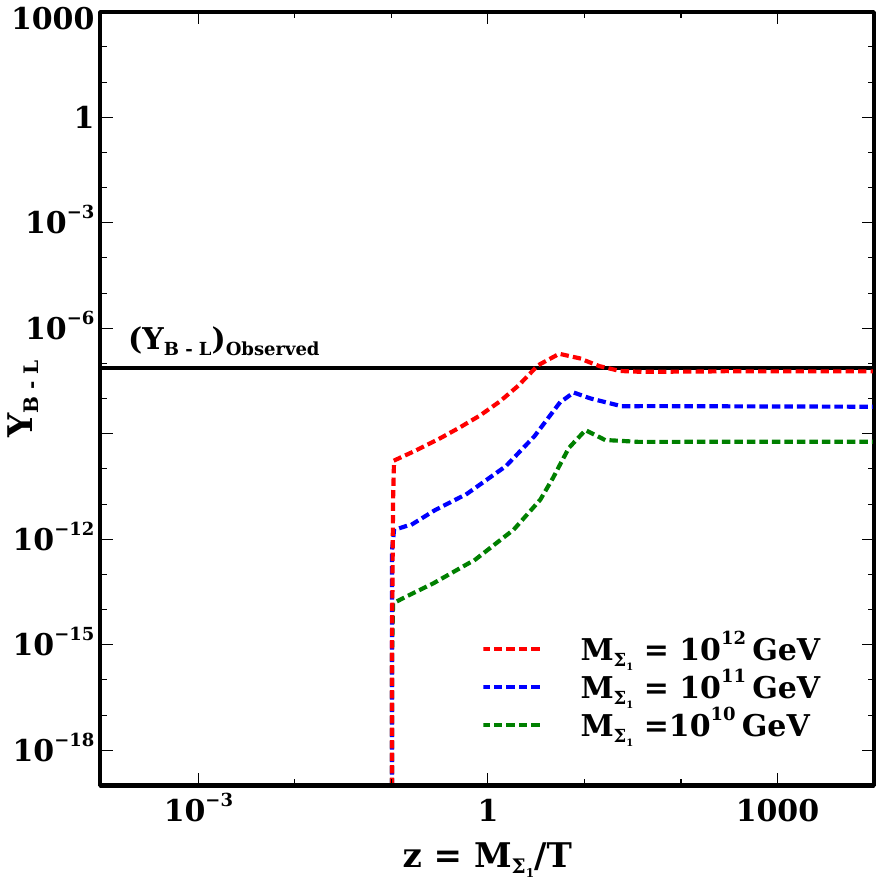}
    \caption{Evolution plot of co-moving number density of $\Sigma_{1}$ (left panel) and $B - L$ (right panel) for different values of mass of fermion triplet $M_{\Sigma_{1}}$. The mass difference between two fermion triplets is kept as $\Delta M_{21} = 10$ GeV.}
    \label{fig:standard}
\end{figure}

In Fig. (\ref{fig:standard}) we show the solutions of the Boltzmann equations Eq. (\ref{Boltz1}). We show the variation of the co-moving number density of $\Sigma_{1}$ and $\rm B-L$ asymmetry with the mass of the triplet $M_{\Sigma_{1}}$ keeping other parameters fixed. In the left panel plot of 
Fig. (\ref{fig:standard}) it is observed that irrespective of the mass of the triplet fermion $\Sigma_{1}$, it's abundance remain very close to the equilibrium abundance upto a very low temperature. This is due to the presence of strong gauge annihilations for $\Sigma_{1}$. Although the intercation rate of the gauge annihilation should decrease with the increase in mass $M_{\Sigma_{1}}$, the annihilations are still sufficiently strong
to keep it's abundance close to the equilibrium abundance and any small change is not visible in this plot. On the right panel plot of Fig. (\ref{fig:standard}), one can see that with the increase in $M_{\Sigma_{1}}$, the $\rm B-L$ asymmetry production increases. This happens mainly due to three reasons, (i) with the increase in $M_{\Sigma_1}$, the CP asymmetry $\epsilon_{1}$ increases, (ii) with the increase in $M_{\Sigma_{1}}$, the Yukawa couplings $Y_{\Sigma}$ increase resulting in the increase in decay as well as inverse decay rates of $\Sigma_{1}$ and (iii) with the increase in $M_{\Sigma_{1}}$ the gauge annihilation rate $S_{A}$ decreases enhancing the out-of-equilibrium condition. It is seen that in a standard radiation dominated Universe the $\Sigma_{1}$ has to be as heavy as $10^{12}$ GeV.

\section{Leptogenesis in a Fast Expanding Universe (FEU) }\label{section3}

To study the effect of non-standard cosmology in triplet leptogenesis, we consider the Universe to be dominated by a scalar field $\phi$ and it's energy density to falls faster than radiation. The energy density of such a field can be considered to be $\rho_{\phi} \propto a^{-(4+n)}$, where $n> 0$. Examples of such theories with $n= 2$ are quintessence fluids \cite{Caldwell:1997ii,Sahni:1999gb}. The energy density of such a fluid redshift as $\rho_{\phi}\propto a^{-6}$, in the kination regime, when the kinetic energy density dominates over the
potential energy \cite{Ratra:1987rm, Wetterich:1987fm}. For the continuity equation the energy density of any species can be calculated to be $\rho_{\phi}\propto a^{-3(1+\omega)}$, where $\omega$ is the equation of state parameter. Therefore, one can identify the parameter $n=3\omega-1$. For a faster expansion than kination ($n=2$), the equation of state parameter has to be $\omega\geq 1$. For a normal fluid $\omega\geq 1$ leads to causality violation and therefore not possible. However, in a  ekpyrotic scenario one can have $\omega\geq 1$ ($n\geq2$) \cite{Khoury:2001wf, Choi:1999xn}. To have a significant effect of fast expansion in this work we take $n\leq4$.

Such a non-standard cosmological history of the Universe was first discussed in \cite{DEramo:2017gpl,DEramo:2017ecx}, in the context of weakly interacting massive particle (WIMP) dark matter and feebly interacting massive particle (FIMP) dark matter productions. The effects of such a FEU in leptogenesis is first studied in \cite{Mahanta:2019sfo,Chen:2019etb}, in the scotogenic model and in the type-I seesaw model respectively. In \cite{Biswas:2023azl}, a triplet fermion leptogenesis in a FEU is studied. It is show that with $n\lesssim 1.8$, the triplet fermion leptogenesis scale can be lowered by two orders from the standard cosmology. In \cite{Mahanta:2022gsi}, the authors  studied a wimpy leptogenesis scenario in such a FEU.

To write the modified Boltzmann equation in a FEU we start writing the total energy density of FEU. It can have contribution from the field $\phi$ as well as from standard radiation

\begin{eqnarray}
    \rho(T)=\rho_{\phi}(T)+\rho_{rad}(T).
\end{eqnarray}

\noindent The standard radiation energy density $\rho_{rad}(T) $ is given by

\begin{eqnarray}
    \rho_{rad}(T)= \dfrac{\pi^{2}}{30}g_{*}(T)T^{4},
\end{eqnarray}

\noindent where $g_{*}$ is the number of relativistic degrees of freedom in the Universe. We consider the field $\phi$ to contribute to the energy density of the Universe but not to the entropy density of the Universe. This is ensured by not considering any interaction of the field $\phi$ with the SM particles. In such a case the entropy of the Universe in a comoving volume $S=sa^{3}=\rm constant$ and the entropy density of the Universe can be written as 

\begin{eqnarray}
\label{eq:entropy}
    s(T)= \dfrac{2\pi^{2}}{45}g_{*s}(T)T^{3},
\end{eqnarray}

\noindent where $g_{*s}$ is the relativistic degrees of freedom contributing to the entropy density of the Universe. 

Since the energy density of the scalar field $\phi$ falls faster than radiation, at some point in the early Universe it falls below the energy density of radiation. The temperature ($T_r$) at which $\rho_{\phi}(T)$ and $\rho_{rad}(T)$ become equal must be above the BBN temperature $T_r\gtrsim (15.4)^{1/n}$ MeV. From Eq. (\ref{eq:entropy})
and scaling the equation $\rho_{\phi}\propto a^{-(4+n)}$ one can write $\rho_{\phi}$ in terms of temperature as

\begin{eqnarray}
    \rho_{\phi}(T)=\rho_{\phi}(T_{r}) \left(\frac{g_{*s}(T)}{g_{*s}(T_r)} \right)^{\frac{4+n}{3}}
\left( \frac{T}{T_r} \right)^n.
\end{eqnarray}

\noindent Therefore total energy density of the Universe is 

\begin{align}
\rho(T)
&= \rho_{\mathrm{rad}}(T) + \rho_{\phi}(T), \nonumber \\
&= \rho_{\mathrm{rad}}(T)
\left[
1 + \frac{g_*(T_r)}{g_*(T)}
\left( \frac{g_{*s}(T)}{g_{*s}(T_r)} \right)^{\frac{4+n}{3}}
\left( \frac{T}{T_r} \right)^n
\right].
\end{align}

\noindent Considering $g_{*}=g_{*s}$ for most of the history of early Universe the Hubble expansion rate in FEU can be written as

\begin{eqnarray}
    \textbf{H}^{'}(T) & = & \sqrt{\frac{\rho(T)}{3M_{Pl}^{2}}} \nonumber \\
    & = & \frac{\pi\, g_*^{1/2}(T)\, T^2}{3\sqrt{10}\, M_{\mathrm{Pl}}}
\left[
1 + \left(
\frac{g_*(T)}{g_*(T_r)}
\right)^{\frac{1+n}{3}}
\left(
\frac{T}{T_r}
\right)^n
\right]^{1/2}.
\end{eqnarray}

\noindent In presence of such a scalar field in the early Universe the modified Boltzmann equations for leptogenesis can be written as 

\begin{align}\label{Boltz2}
 \nonumber \frac{dY_{\Sigma_{1}}}{dz} &= -D'_{\Sigma_{1}} (Y_{\Sigma_{1}}-Y^{eq}_{\Sigma_{1}})-S'_{A}(Y_{\Sigma_{1}}^{2}-(Y_{\Sigma_{1}}^{eq})^{2}), \\
  \frac{dY_{B-L}}{dz} &= -\epsilon_{\Sigma_{1}} D'_{\Sigma_{1}} (Y_{\Sigma_{1}}-Y^{eq}_{\Sigma_{1}}) - W_{\Sigma_1}^{'} Y_{B-L} - W'_{\Delta L} Y_{B-L}.
\end{align}

\noindent The decay terms $D'_{\Sigma_{1}}$ is given by 
\begin{align}
  D'_{\Sigma_1} = K_{\Sigma_1} \left( z \right) \frac{\kappa_{1}\left(z\right)}{\kappa_{2}\left(z\right)}\frac{1}{L[n, z, z_r]}  . 
\end{align}

\noindent The annihilation term $S_{A}^{'}$ for $\Sigma_{1}$ gets modified as

\begin{equation}
    S'_{A} = \bigg(\frac{\pi^2 g^{*1/2}M_Pl}{1.66*180 g_{\Sigma_1}^2}\bigg)\frac{1}{M_{\Sigma_1}}\bigg(\frac{I_{z}}{ \kappa_{2}(z)^{2}}\bigg)\frac{1}{ (z)^2L[n, z, z_{r}]}.
\end{equation}

\noindent The modified inverse decay term $W_{\Sigma_{1}}^{'}$

 \begin{align}
     W_{\Sigma_1}^{'} = \frac{1}{4}K_{\Sigma_1}( z)^{3}\kappa_{1}\left(z\right)\frac{1}{L[n, z, z{_r}]}.
 \end{align}

\noindent Similarly, the scattering washout term also modifies and is given by
 
\begin{align}
W'_{\Delta L}=\frac{\Gamma_{scattering}}{\textbf{H} z^{2}L[n, z, z{_r}]}.    
\end{align}

\noindent The function $L[n,z,z_r]$, that modifies the interaction terms in the Boltzmann equations is given by 
\begin{equation}
    L[n,z,z_r] = (n+4)\Bigg[\frac{1}{z^{4}}+\bigg(\frac{g_{*}(z)}{g_{*}(z_r)}\bigg)^{(1+n)/3}\frac{z_{r}^{n}}{z^{n+4}}\Bigg]^{3/2}\bigg[\frac{4}{z^{5}}+(4+n)\bigg(\frac{g_{*}(z)}{g_{*}(z_r)}\bigg)^{(1+n)/3}\frac{z_{r}^{n}}{z^{n+5}}\bigg]^{-1}.
\end{equation}

\begin{figure}[h]
    \centering
    \includegraphics[scale=0.45]{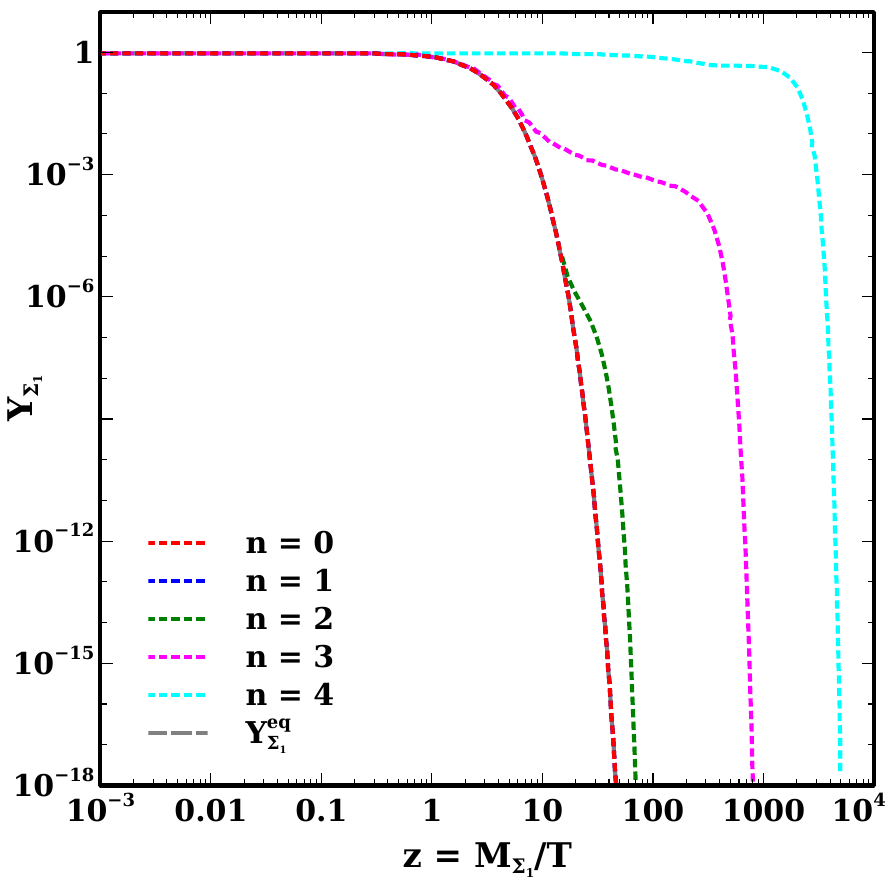}
\includegraphics[scale=0.45]{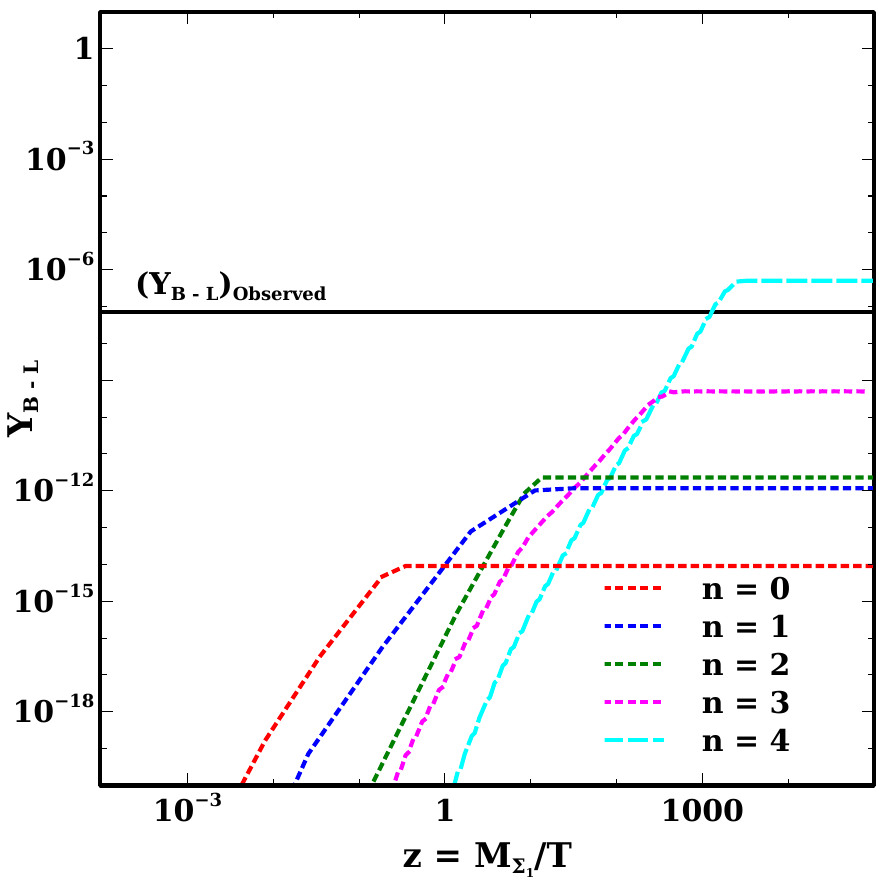}
    \caption{Evolution plot of the co-moving number density of $\Sigma_{1}$ (left panel)and $\rm B-L$ (right panel) for different values of n. The parameters used are $M_{\Sigma_1} = 10$ TeV and $T_{r} = 0.1$ MeV. The mass difference between $\Sigma_{1}$ and $\Sigma_{2}$ is given by $\Delta M_{21} = 10$ GeV.}
    \label{fig:FEU1}
\end{figure}

\begin{figure}[h]
    \centering
    \includegraphics[scale=0.45]{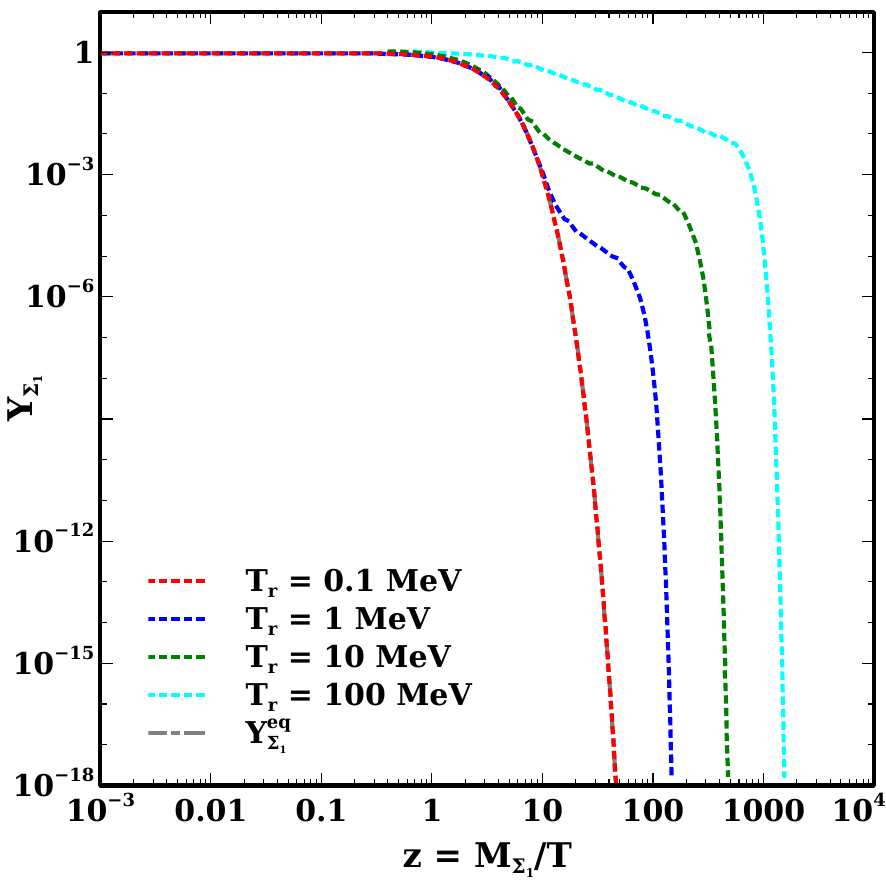}
    \includegraphics[scale=0.45]{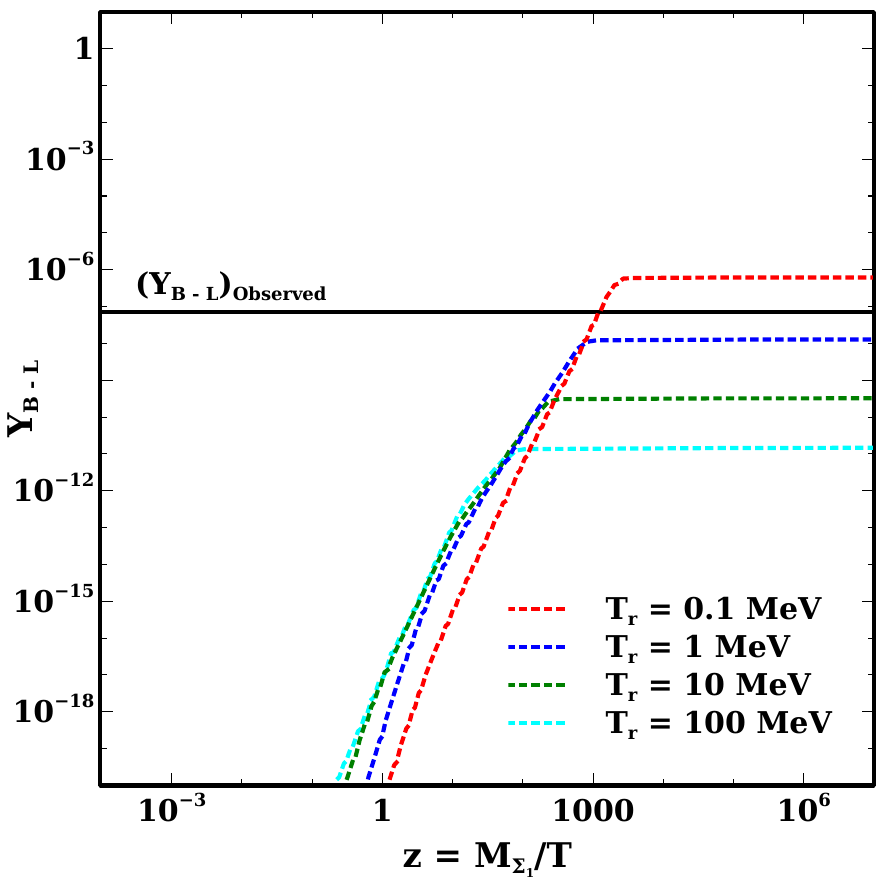}
    \caption{Evolution plot of the co-moving number density of $\Sigma_1$ (left panel) and $\rm B-L$ asymmetry (right panel) for different values of $T_{r}$ keeping other parameters fixed. The parameters used are $n=4$ and $M_{\Sigma_1}=10$ TeV. The mass difference between $\Sigma_1$ and $\Sigma_2$ is given by $\Delta M_{21}=100$ GeV. }
    \label{fig:FEU2}
\end{figure}

\begin{figure}[h]
    \centering
    \includegraphics[scale=0.45]{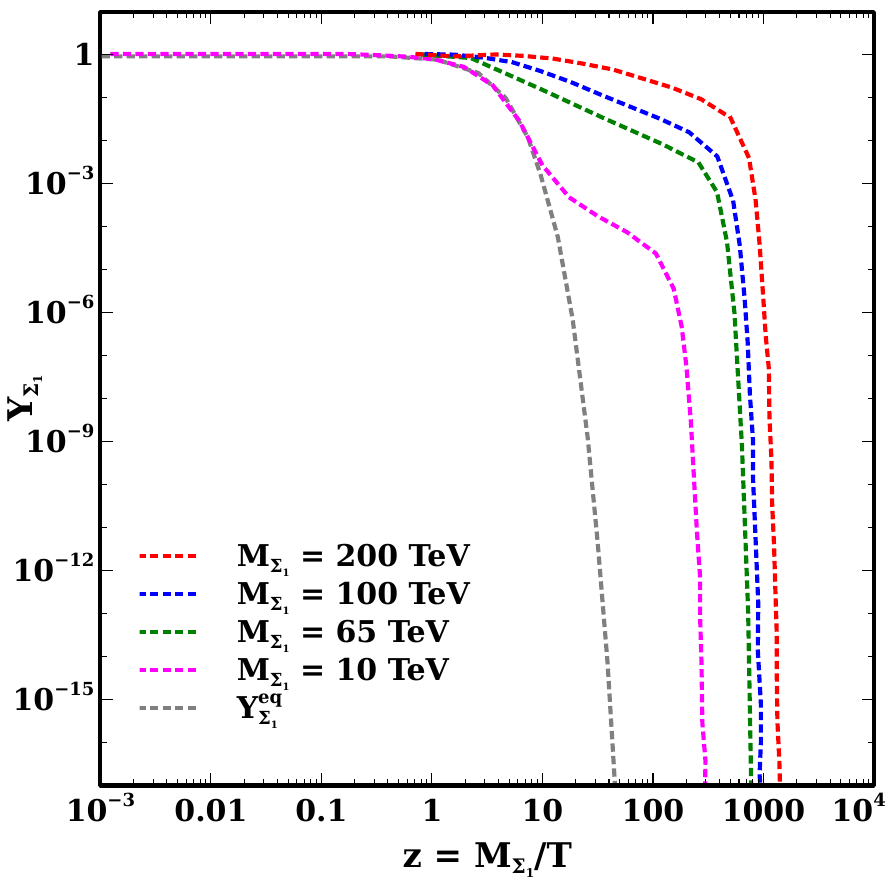}
    \includegraphics[scale=0.45]{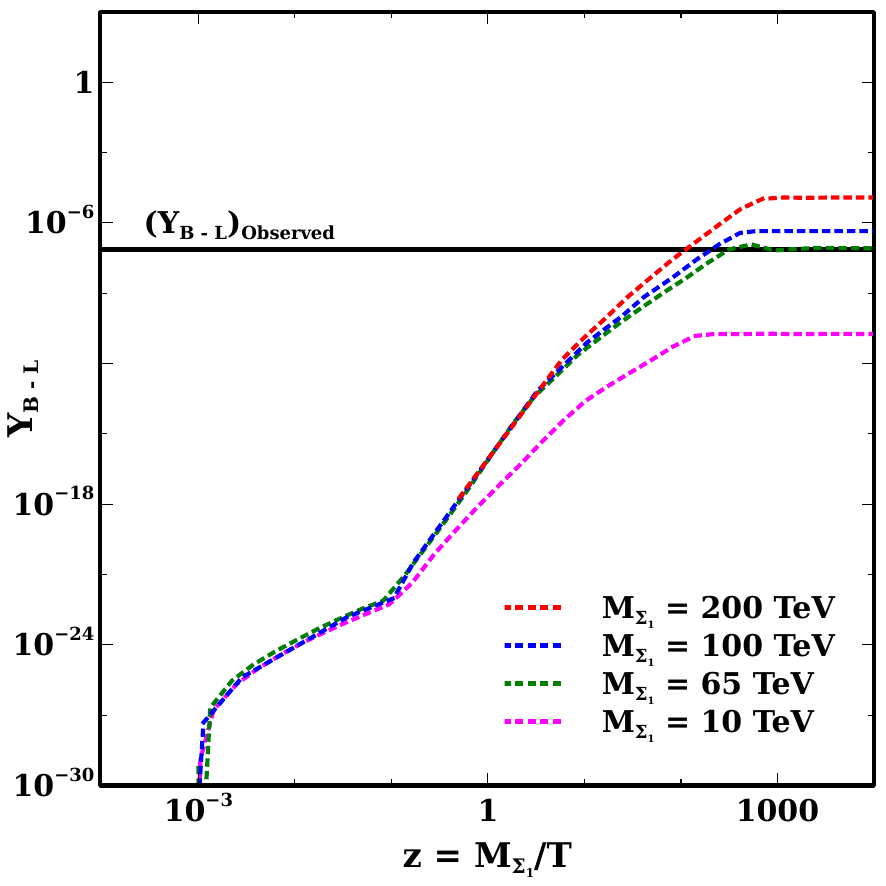}
    \caption{Evolution plot of the co-moving number density of $\Sigma_1$ (left panel) and $\rm B-L$ asymmetry (right panel) for different values of $M_{\Sigma_1}$ keeping other parameters fixed. The cosmological parameters are fixed at $n=4$ and $T_r=0.1 \rm MeV$. The mass difference between $\Sigma_1$ and $\Sigma_2$ is given by $\Delta M_{21}=100$ GeV.}
    \label{fig:FEU3}
\end{figure}

\begin{figure}[h]
    \centering
    \includegraphics[scale=0.45]{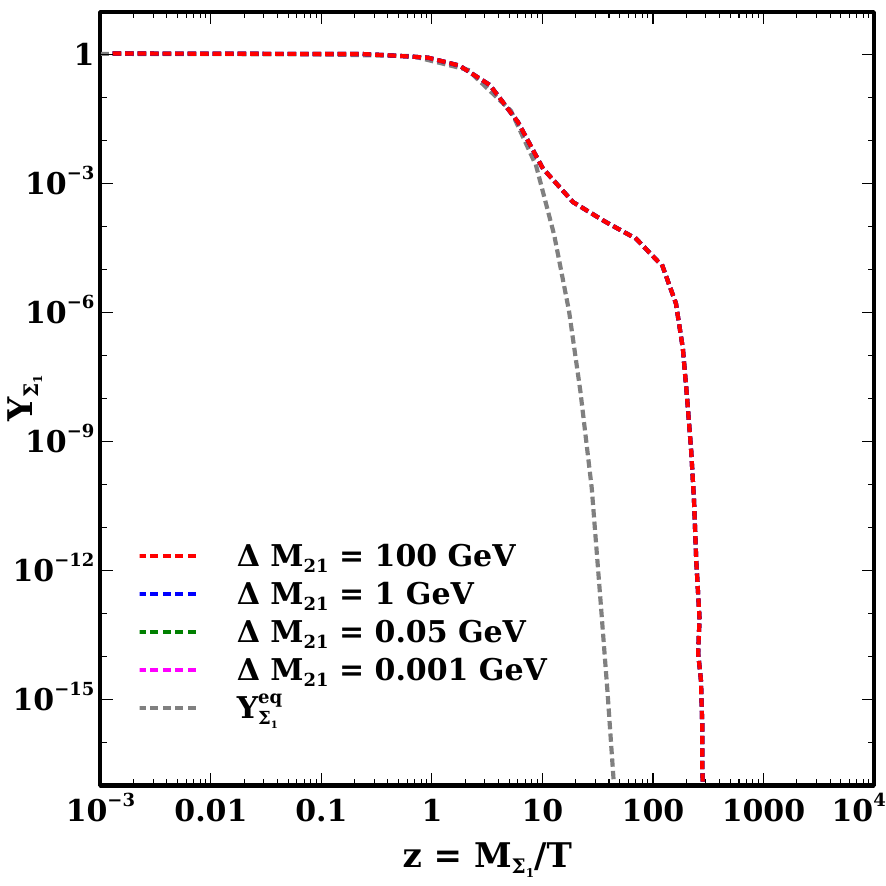}
    \includegraphics[scale=0.45]{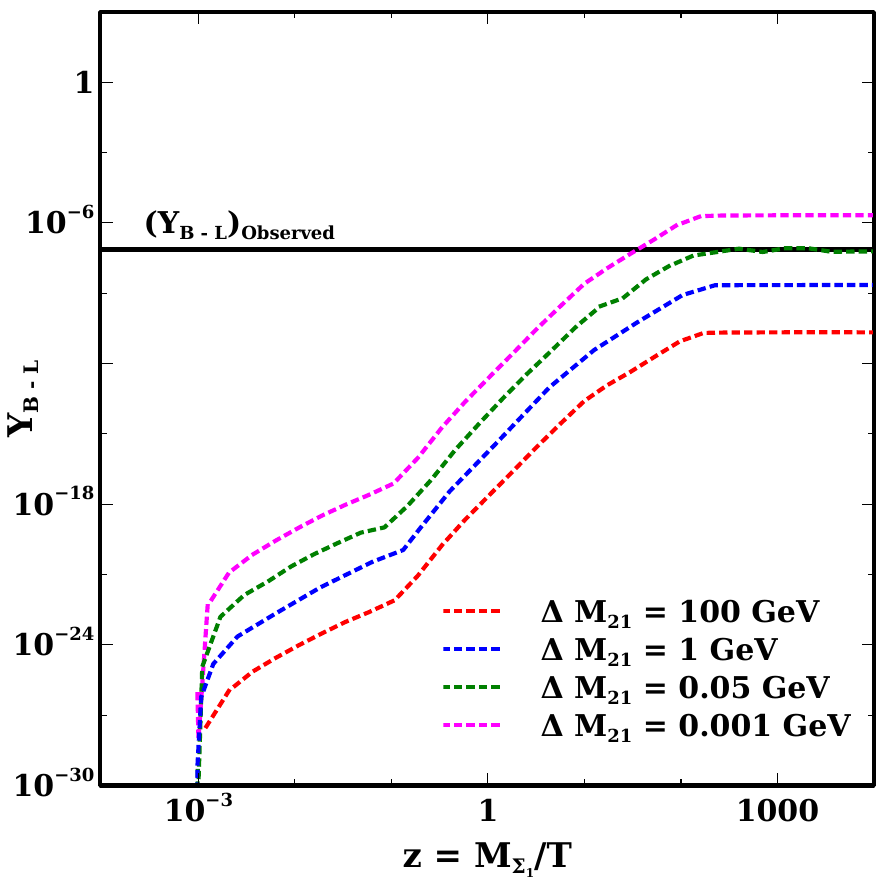}
    \caption{Evolution plot of the co-moving number density of $\Sigma_1$ (left panel) and $\rm B-L$ asymmetry (right panel) for different values of $\Delta M_{12}$ keeping other parameters fixed. The cosmological parameters are fixed at $n=4$ and $T_r=0.1 \rm MeV$. The mass of $\Sigma_1$ is given by $M_{\Sigma_1}=10$ TeV. }
    \label{fig:FEU4}
\end{figure}

\begin{figure}[h]
    \centering
\includegraphics[scale=0.45]{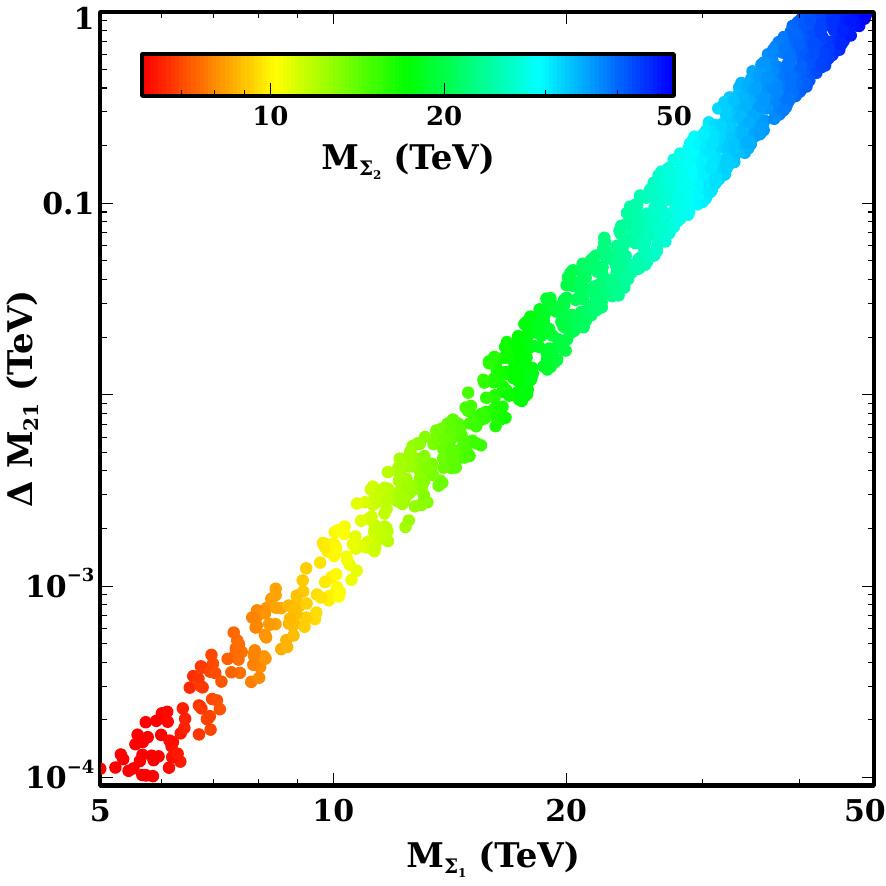}
    \caption{Scan plot in $M_{\Sigma_1}$ vs $\Delta M_{21}$ plane showing the parameter space that generate the observed baryon asymmetry. The mass of $\Sigma_2$ is shown in the colorbar. The cosmological parameters are fixed at $n=4$ and $T_r=0.1 \rm MeV$.}
    \label{fig:FEU_scan}
\end{figure}

We show the solution of the Boltzmann equations Eq. (\ref{Boltz2}) in Fig. (\ref{fig:FEU1}) and Fig. (\ref{fig:FEU2}). In Fig. 
(\ref{fig:FEU1}) we show evolution of the co-moving number density $Y_{\Sigma_1}$ and $Y_{B-L}$ with $z=M_{\Sigma_1}/T$ for different benchmark values of $n$ keeping other parameters fixed. From the left panel plot of Fig. (\ref{fig:FEU1}) it is observed that a faster expansion (larger $n$) takes $\Sigma_1$ away from its equilibrium abundance earlier. It results in an increase in the production of asymmetry as well as a decrease in the washout effects. In the Fig. (\ref{fig:FEU2}) we show the evolution of the co-moving number density of $\Sigma_1$ and $\rm B-L$ asymmetry for different values of $T_r$. From the left panel plot it can be observed that for a small value of $T_r$, the Universe expand faster than radiation for a long period resulting in larger deviation of $\Sigma_1$ from its equilibrium abundance. In the plot on the right panel it can be seen that a smaller $T_r$ increases the production of asymmetry. In Fig. (\ref{fig:FEU3}) we show the evolution plots for the number density of $\Sigma_1$ and $\rm B-L$ asymmetry for different benchmark values of $M_{\Sigma_{1}}$. It can be seen that with an increase in $M_{\Sigma_1}$, $\Sigma_1$ deviates from equilibrium earlier resulting an increase in the asymmetry production. It can be seen that the observed asymmetry can be generated with $M_{\Sigma_1} \simeq 65$ TeV with $\Delta M_{21}=M_{\Sigma_{2}}-M_{\Sigma_{1}}=100$ GeV by appropriately choosing the FEU parameters $T_{r}$ and $n$. In Fig. (\ref{fig:FEU4}) we show the evolution of the co-moving number density of $\Sigma_1$ and $\rm B-L$ asymmetry with different values of $\Delta M_{21}$ keeping other important parameters at fixed benchmark values. We keep the cosmological parameters $n=4$ and $T_r=0.1$ MeV to maximize the asymmetry. In the right panel plot of Fig. (\ref{fig:FEU4}) it is seen that it is possible to generate the observed asymmetry with $M_{\Sigma_1}\simeq 10$ TeV keeping the mass difference sufficiently small ($\Delta M_{21}\simeq 0.01 \rm GeV)$. It is important to note that although we require a small mass difference between $\Sigma_2$ and $\Sigma_1$ to have successful leptogenesis the TeV scale we are still away from the resonant condition $\Delta M_{21}\simeq \Gamma_{2,1} \simeq \mathcal{O}(10^{-4}\rm GeV)$. In Fig. (\ref{fig:FEU_scan}) we show the parameter space in $M_{\Sigma_1}-\Delta M_{21}$ plane with $M_{\Sigma_2}$ as colorbar. Here we keep the cosmological parameters at $n=4$ and $T_r=0.1$ MeV to maximize the effect of fast expansion. Since a small value of $\Delta M_{21}$ enhances the CP asymmetry through the self energy contribution, a small value of $M_{\Sigma_1}$ is require a smaller $\Delta M_{21}$ to satisfy the correct $\rm B-L$ asymmetry.

\section{Leptogenesis in Scalar Tensor Theory of Gravity (STG)} \label{section4}

In this section we study the impact of modified expansion rate of the Universe due to a class of Scalar-Tensor theory of gravity (STG) on triplet leptogenesis \cite{Fierz:1956zz, Brans:1961sx}. In STG, gravity is described not only by geometry but also by a scalar field. In such theories, the rate of cosmic expansion deviates from that of standard general relativity (GR). An attractor mechanism relaxes it to the standard expansion era prior to the BBN era. STG's are generally formulated in an Einstein frame or in Jordan frame. The most general transformation between the metric written in Jordan frame $\tilde{g}_{\mu \nu}$ and Einstein frame $g_{\mu \nu}$ is given by 

\begin{eqnarray}
    \tilde{g}_{\mu \nu}= C(\phi)g_{\mu \nu}+D(\phi)\partial_{\mu}\phi\partial_{\nu}\phi .
\end{eqnarray}

\noindent Here, $C(\phi)$ and $D(\phi)$ are known as conformal and disformal couplings,, respectively. In Jordan frames, the matter fields $\Psi_{i}$ directly couple to the metric $\title{g}_{\mu \nu}$
and the action can be written as $S_{Matter}=S_{Matter}(\tilde{g}_{\mu \nu}, \Psi_i)$. In such a case the effect of modified gravity changes the expansion rate of the Universe while the particle physics observables remain unchanged. On the other hand in the Einstein frame the scenario becomes completely opposite. Therefore, we chose to work in the Jordan frame throughout this work. In \cite{Dutta:2018zkg}, the authors have studied Leptogenesis in such a STG theory, while the effect of such non-standard cosmology in case of DM relic were studied in several other works  \cite{Catena:2004ba, Catena:2009tm,Meehan:2015cna, Dutta:2017fcn, Dutta:2016htz }. 

We follow the procedure given in \cite{Dutta:2016htz, Dutta:2017fcn} and write the master equation to track the evolution of the scalar field $\phi$ in the conformal limit ($D(\phi)=0$)

\begin{eqnarray}
    \frac{2}{3B[1-\alpha(\varphi)\varphi']^3}\bigg(\varphi''+\frac{d\alpha}{d\varphi}(\varphi')^3\bigg) + \frac{1-\tilde{\omega}}{[1-\alpha(\varphi)\varphi']}\varphi' + 2(1-3\tilde{\omega})\alpha(\varphi) = 0.
    \label{eq:master}
\end{eqnarray}

Here $\varphi = \kappa\phi$ is a dimensionless scalar introduced for convenience and $B = 1-\frac{1}{6}\frac{\varphi'}{(1-\alpha(\varphi)\varphi')^2}$. The function $\alpha(\varphi) = \frac{d \rm ln C^{1/2}}{d\varphi}$, where the conformal coupling $C(\varphi)$ is considered to be $C(\varphi) = (1 + 0.1 exp(-8\varphi))^{2}$. The choice of such conformal coupling is motivated by the earlier works \cite{Catena:2004ba, Dutta:2016htz}. Here the derivatives are taken with respect to the number of e-folds $d\tilde{N}=\tilde{H}dt$ in Jordan frame. The $\tilde{\textbf{H}}$ is the modified Hubble parameter in STG theory. The number of e-folds can be written as Jordan frame temperature as follows

\begin{eqnarray}
    \tilde{N}=\rm ln \left[  \frac{\tilde{T_{0}}}{\tilde{T}}\left( \frac{g_{*s}(\tilde{T_0})}{g_{*s}(\tilde{T})} \right)^{1/3} \right]. 
\end{eqnarray}

\noindent The Hubble expansion rate is modified and is given by

\begin{eqnarray}
    \tilde{\textbf{H}}^{2}= \dfrac{k^{2}}{k_{GR}^{2}}\dfrac{C(1+\alpha(\phi)\phi^{'})^{2}}{1-\varphi^{'2}/6}\textbf{H}^{2}. 
    \label{eq:Hub1}
\end{eqnarray}

\noindent Where $\textbf{H}^{2}=\dfrac{k_{GR}^{2}}{3}\tilde{\rho}$, $k^{2}\approx k^{2}_{GR}=8 \pi G$ and $\tilde{\rho} \sim g(\tilde{T})\tilde{T}^{4}$. From Eq. (\ref{eq:Hub1}) we define the speed up parameter $\xi$, as the ratio of modified Hubble expansion rate by the standard radiation dominated Hubble expansion rate

\begin{eqnarray}
    \xi= \dfrac{\tilde{\textbf{H}}}{\textbf{H}}=\left[  \dfrac{k^{2}}{k^{2}_{GR}} \dfrac{C(1+\alpha(\phi)\phi^{'})^{2}}{1-\varphi^{'2}/6}\right]^{1/2}.
\end{eqnarray}

The third term in Eq. (\ref{eq:master}) behaves like an effective potential $V_{eff}=\rm ln C^{1/2}$. In the radiation dominated Universe the effective potential vanishes as $\tilde{\omega}=1/3$. Later when the Universe cools down and the particles start becoming non-relativistic the, $\tilde{\omega}$ starts deviating away from $1/3$. As a result the effective potential kicks in. The solution of the master equation in the absence of any effective potential reads $\varphi^{'} \propto e^{-\tilde{N}}$. This means any initial value of the velocity $\varphi^{'}$ will instantly become zero. We choose a positive initial value of $\varphi$ and a negative value of $\varphi^{'}$ as discussed in \cite{Dutta:2016htz,Dutta:2018zkg}. This choice leads to a very interesting scenario in which the field $\varphi$ moves to a negative value until its velocity becomes zero and then becomes positive as the field moves down the effective potential. This change in the evolution of the scalar field leads to a peak in the conformal coupling $C(\varphi)$, that results in a peak in the modified expansion rate of the Universe $\tilde{H}$ in the Jordan frame. Later as the field rolls down to positive values the conformal coupling becomes one retrieving the standard GR expansion rate of the Universe.

To calculate the equation of state parameter $\tilde{\omega}$ during the early stage of the Universe we write

\begin{equation}
\centering
\label{eq:omega}
    1-3\tilde{\omega}=\dfrac{\tilde{\rho}-3\tilde{p}}{\tilde{\rho}}= \sum_{i}\dfrac{\tilde{\rho}_{i}-3\tilde{p}_{i}}{\tilde{\rho}}+ \dfrac{\tilde{\rho_{m}}}{\tilde{\rho}},
\end{equation}

\noindent where the sum runs over all the particles that enter into the thermal bath during the early Universe. Here $\tilde{\rho}_{m}$ is the energy density of non-relativistic pressure less matter that is negligible during the radiation domination phase. Therefore the Eq. (\ref{eq:omega}) simplifies to

\begin{eqnarray}
    \tilde{ \omega} = \dfrac{1}{3} \left(  1- \sum_{i} \dfrac{\tilde{\rho}_{i}-3\tilde{p}_{i}}{\tilde{\rho}} \right). 
\end{eqnarray}

\begin{figure}
    \centering
    \includegraphics[width=0.4\linewidth]{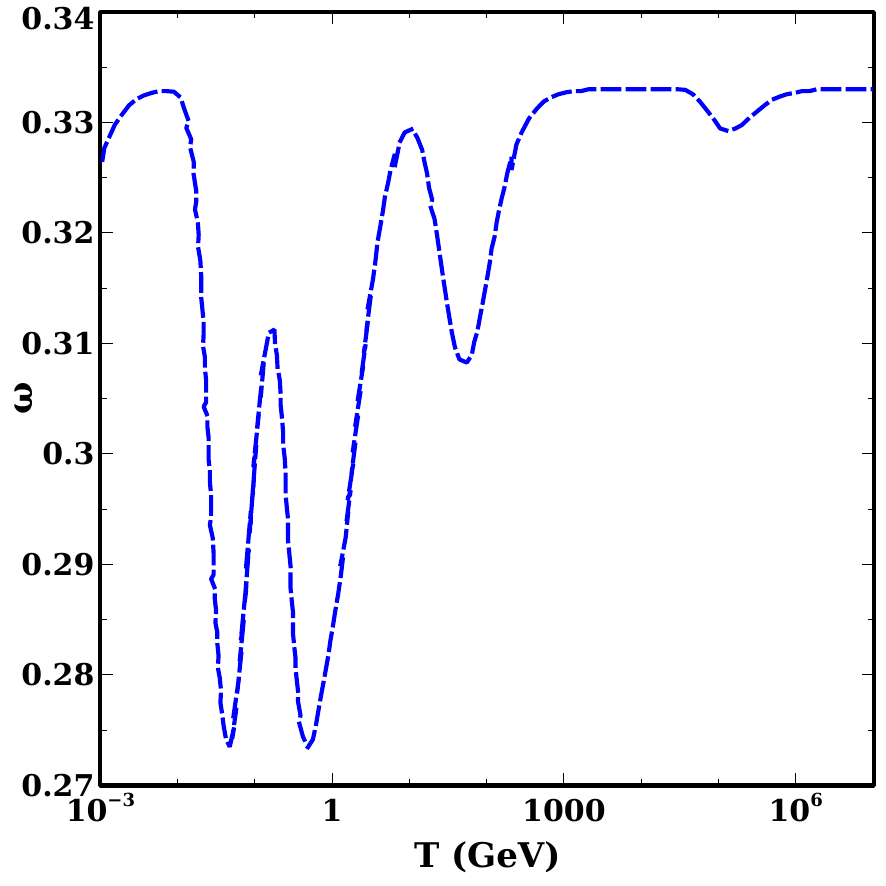}
    \caption{Equation of state parameter as a function of temperature. The mass of $\Sigma_{1}$, $\Sigma_{2}$ are fixed at $M_{\Sigma_{1}} = 350$ TeV, $M_{\Sigma_{2}} = 350.001$ TeV and the cosmological parameters are fixed at $(\varphi)_{0}=0.1$ and $(\varphi')_{0}=-0.79$.}
    \label{fig:omega}
\end{figure}

\begin{figure}[h]
    \centering
    \includegraphics[width=0.4\linewidth]{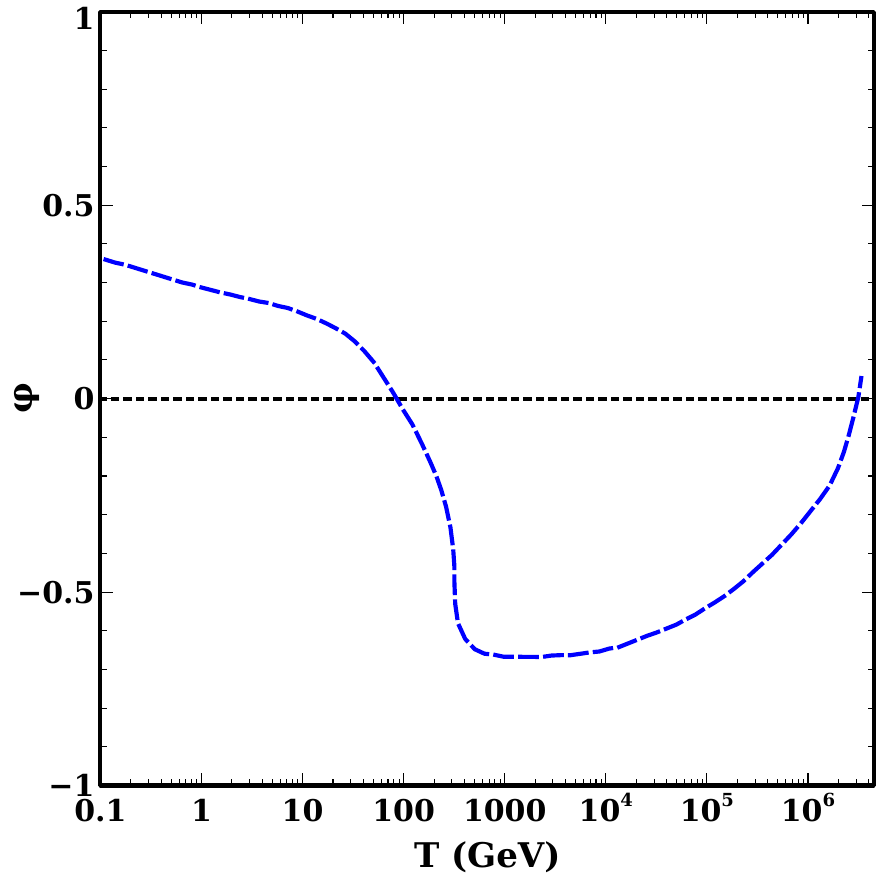}   \includegraphics[width=0.4\linewidth]{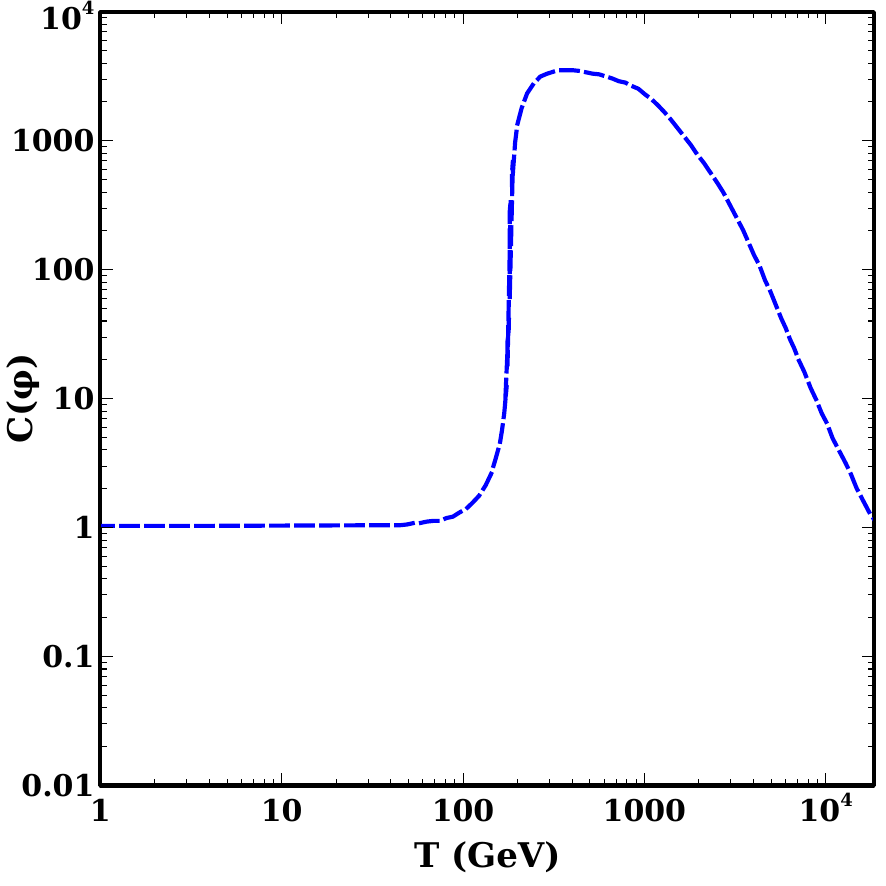}
   
    \includegraphics[width=0.42\linewidth]{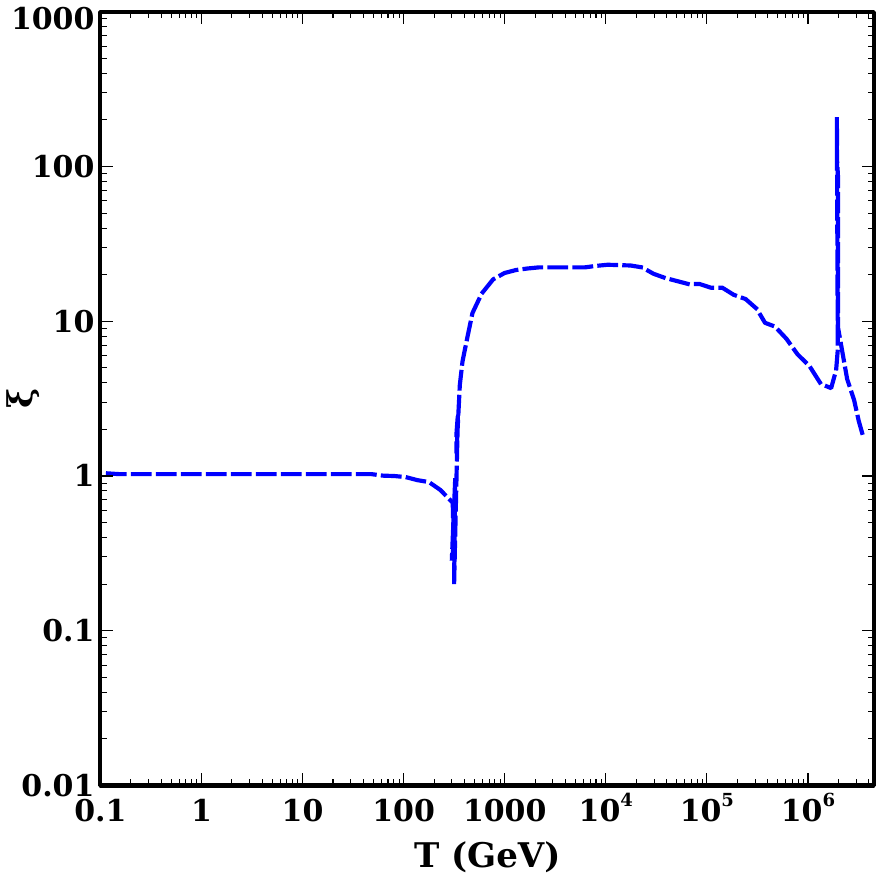}
    \includegraphics[width=0.42\linewidth]{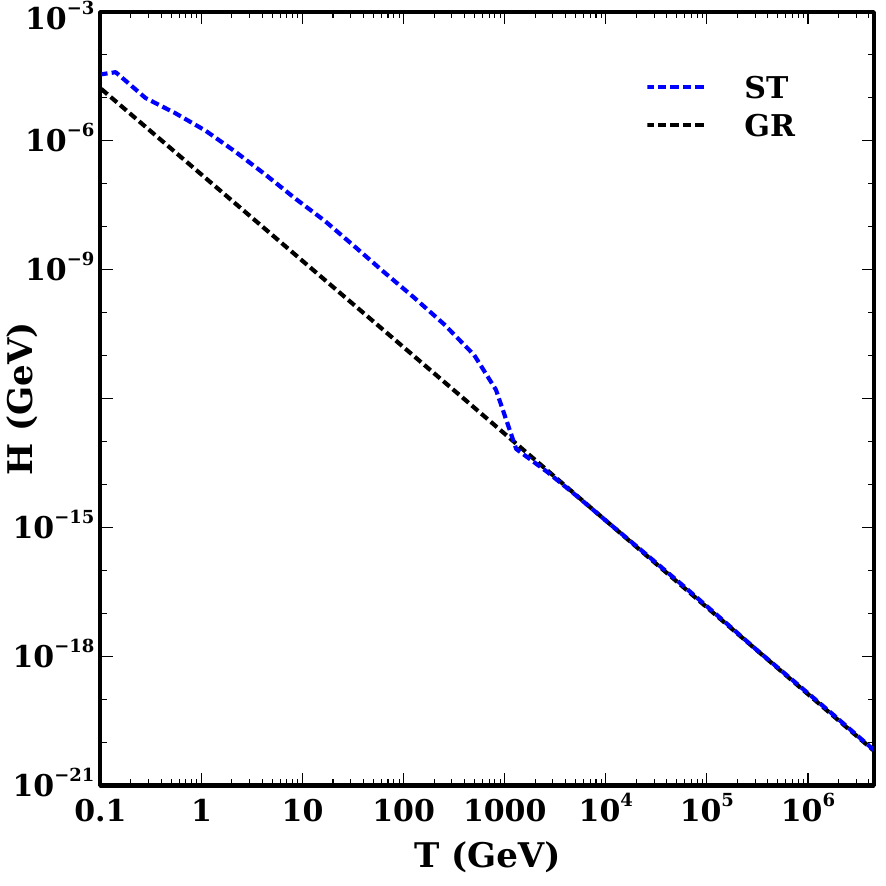}
    \caption{The evolution of field (upper left plot), the conformal coupling (upper right plot), the speed up parameter (lower left plot) and the Hubble expansion rate (lower right plot) with temperature. The mass of $\Sigma_{1}$, $\Sigma_{2}$ are fixed at $M_{\Sigma_{1}} = 350$ TeV, $M_{\Sigma_{2}} = 350.001$ TeV and the cosmological parameters are fixed at $(\varphi)_{0}=0.1$ and $(\varphi')_{0}=-0.79$.}
    \label{fig:field_evolution}
\end{figure}

\noindent The energy density and pressure of any species are calculated by

\begin{eqnarray}
    \tilde{\rho}_{i} & = & \dfrac{g_{i}}{2\pi^{2}}\int_{m_{i}}^{\infty} \dfrac{(E^{2}-m_{i}^{2})^{1/2}}{e^{E/\tilde{T}}\pm 1}E^{2}dE,   \\
    \tilde{p}_{i} & = & \dfrac{g_{i}}{2\pi^{2}}\int_{m_{i}}^{\infty} \dfrac{(E^{2}-m_{i}^{2})^{3/2}}{e^{E/\tilde{T}}\pm 1}dE,
\end{eqnarray}

\noindent where $g_{i}$ are the internal degrees of freedom of the species $i$ and the plus sign corresponds to the boson while the minus sign corresponds to fermions. In the calculation of $\tilde{\omega}$ we consider all the relevant particle in our model. In Fig. (\ref{fig:omega}) we show the evolution of $\tilde{\omega}$ with temperature $\tilde{T}$. As the individual particle become non-relativistic at different temperature the kinks in $\tilde{\omega}$ appear at different temperatures. Solving the master equation Eq. (\ref{eq:master}) we show the evolution of the field $\varphi$, the conformal coupling $C(\varphi)$, and the speed up parameter in Fig. (\ref{fig:field_evolution}). As discussed earlier, the only change is in the Hubble expansion rate that is relevant for leptogenesis. A positive initial value of the scalar field $\varphi$ and a negative initial value of $\varphi^{'}$ gives rise to interesting scenario where the scalar field moves from a positive value to negative value till its velocity becomes zero and becomes
positive again as the field rolls back down the effective potential. The change in the scalar field enhances the conformal coupling as shown in the top right panel plot of Fig. (\ref{fig:field_evolution}) and as a result the Hubble expansion rate pushes up compared to the standard GR expansion rate as shown in Fig. (\ref{fig:field_evolution}).

We then write the modified Boltzmann equations for leptogenesis as in Eq. (\ref{Boltz3}) and solve them with the master equation for $\varphi$

\begin{figure}[h]
        \centering
        \includegraphics[scale=0.45]{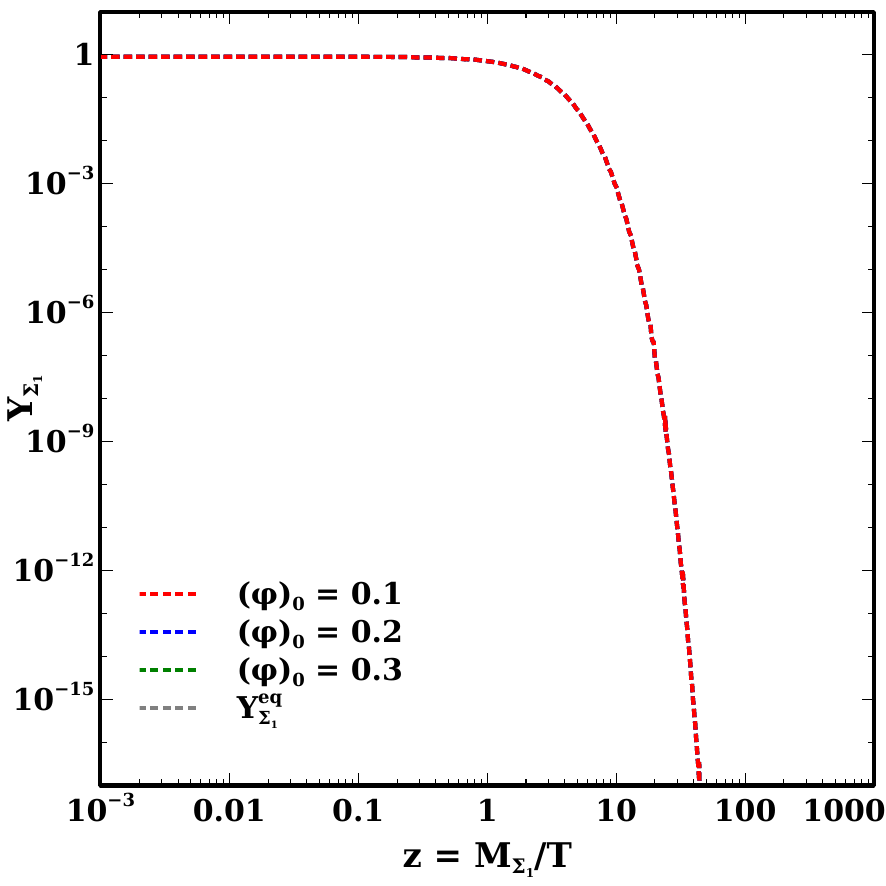}  \includegraphics[scale=0.45]{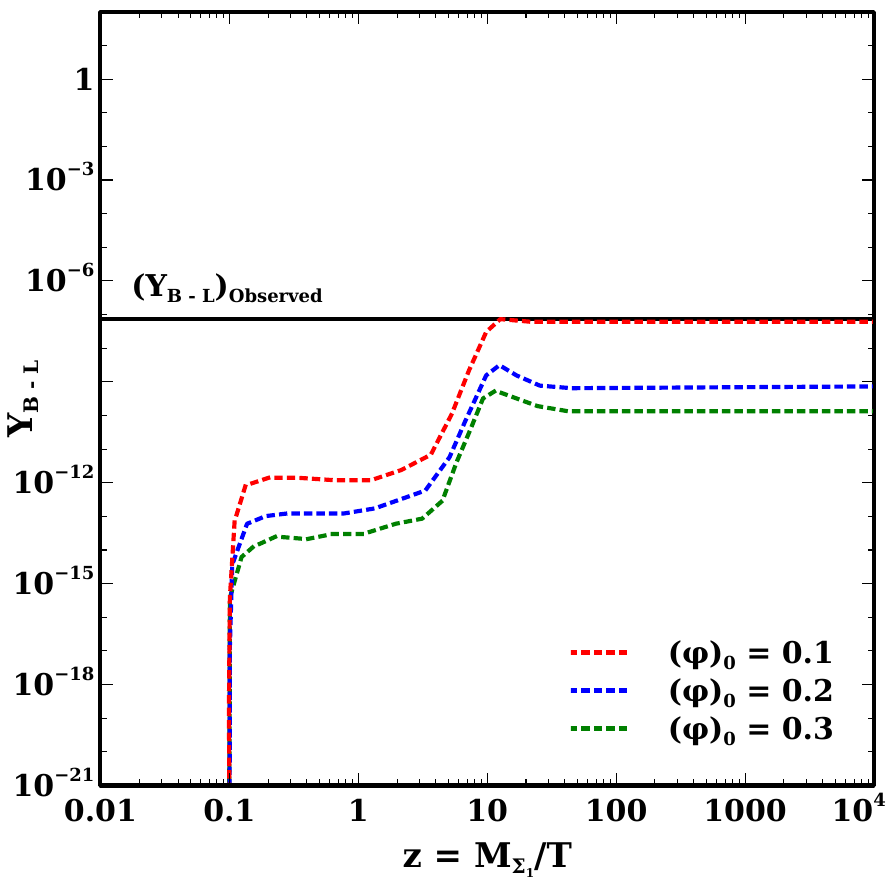}
    \caption{Evolution plot of the co-moving number density of $\Sigma_1$ (left panel) and $\rm B-L$ asymmetry (right panel) for different values of $(\varphi)_{0}$ keeping other parameters fixed. The masses of $\Sigma_{1}$, $\Sigma_{2}$ are fixed at $M_{\Sigma_{1}} = 350$ TeV, $M_{\Sigma_{2}} = 350.001$ TeV and $(\varphi')_{0} = -0.93$. }
    \label{fig:STG1}
\end{figure}

\begin{figure}[h]
\centering
\includegraphics[scale=0.45]{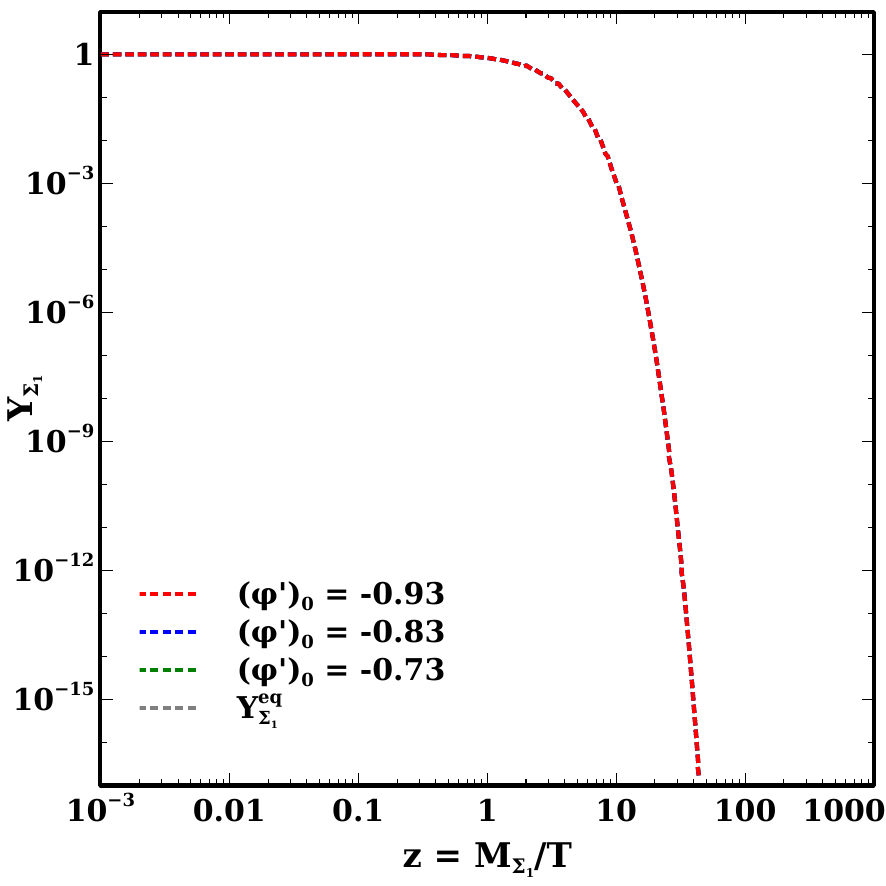}  \includegraphics[scale=0.45]{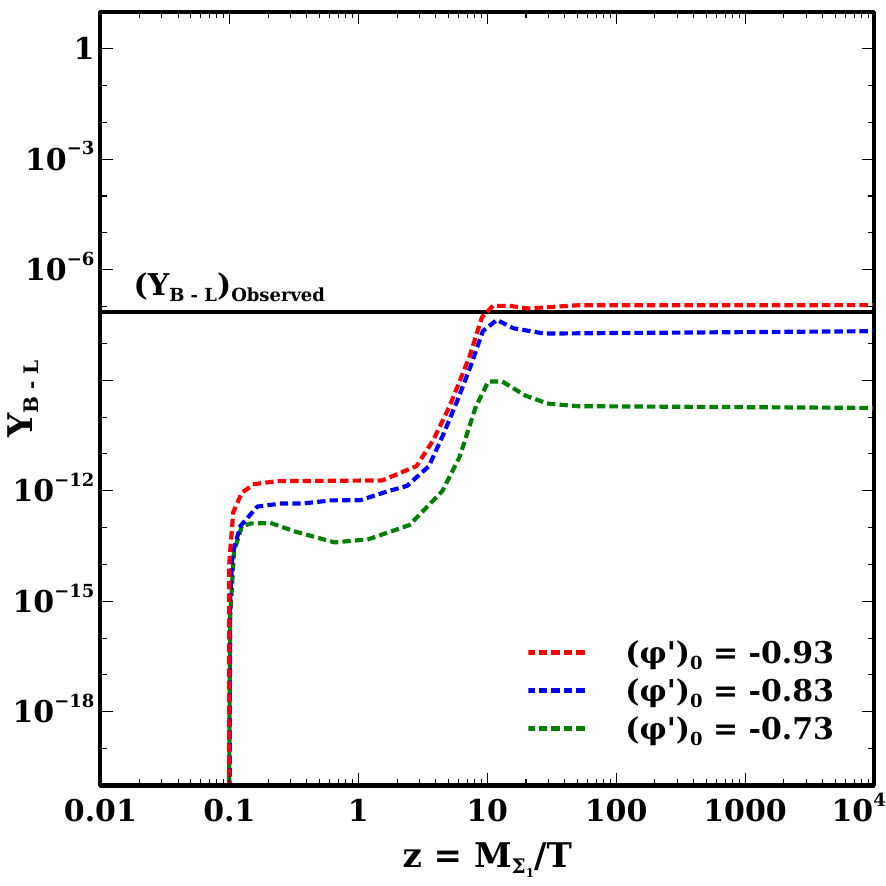}
    \caption{Evolution plot of the co-moving number density of $\Sigma_1$ (left panel) and $\rm B-L$ asymmetry (right panel) for different values of $(\varphi')_{0}$ keeping other parameters fixed. The masses of $\Sigma_{1}$, $\Sigma_{2}$ are fixed at $M_{\Sigma_{1}} = 350$ TeV, $M_{\Sigma_{2}} = 350.001$ TeV and $(\varphi)_{0} = 0.1$.}
    \label{fig:STG2}
\end{figure}

\begin{figure}[h]
    \centering
    \includegraphics[scale=0.45]{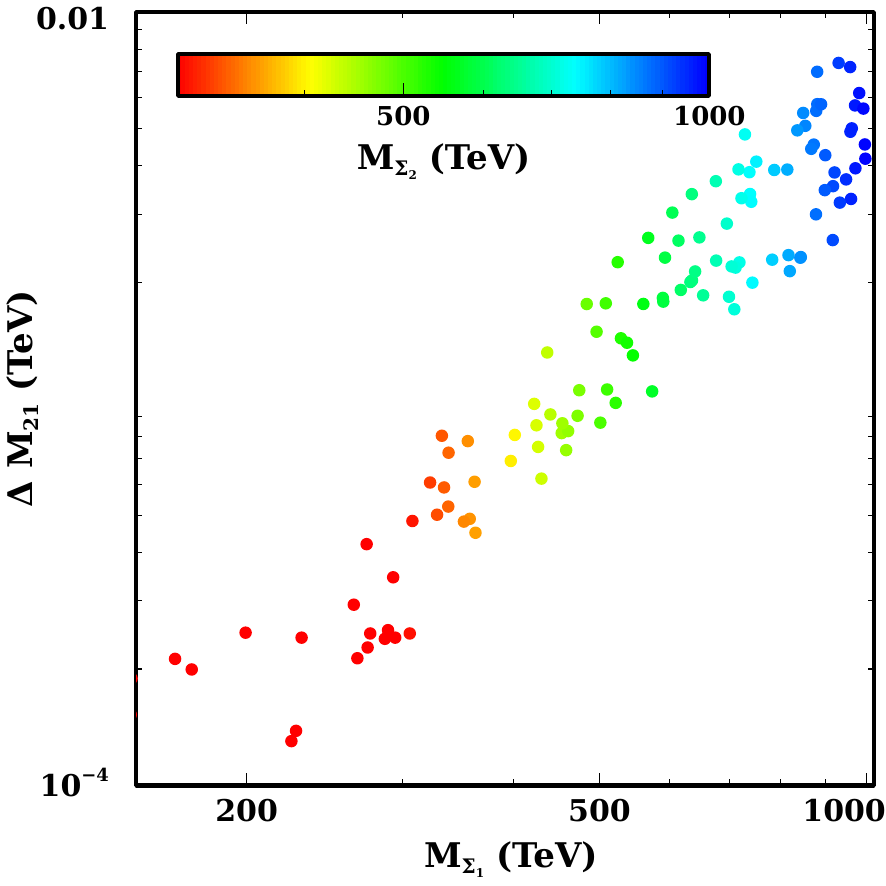}
    \caption{Scan plot in $M_{\Sigma_1}$ vs $\Delta M_{21}$ plane showing the parameter space that generate the observed baryon asymmetry. The mass of $\Sigma_2$ is shown in the colorbar. The cosmological parameters are fixed at $(\varphi)_{0}=0.1$ and $(\varphi')_{0}=-0.93$. }
    \label{fig:STG_scan}
\end{figure}

\begin{align}\label{Boltz3}
\nonumber \frac{dY_{\Sigma_{1}}}{dz} &= -D''_{\Sigma_1} (Y_{\Sigma_{1}}-Y^{eq}_{\Sigma_{1}})-S''_{A}(Y_{\Sigma_{1}}^{2}-(Y_{\Sigma_{1}}^{eq})^{2}),
\\
  \frac{dY_{B-L}}{dz} &=-\epsilon_{\Sigma} D''_{\Sigma_{1}} (Y_{\Sigma_{1}}-Y^{eq}_{\Sigma_{1}}) - W''_{\Sigma_{1}} Y_{B-L}- W''_{\Delta L} Y_{B-L},
\end{align}
The modified decay term $ D''_{\Sigma_{1}}$, scattering term $S''_{A}$, inverse decay term $W''_{\Sigma_{1}}$ and the modified scattering washout term $W''_{\Delta L}$ are given as 

\begin{align}
  D''_{\Sigma_{1}} =\frac{K_{\Sigma_1}}{\xi(z)}  \frac{\kappa_{1}\left(z\right)}{\kappa_{2}\left(z\right)}, 
\end{align}
\begin{equation}
    S''_{A} = \bigg(\frac{\pi^2 g^{*1/2}M_Pl}{1.66*180 g_{\Sigma_{1}}^2}\bigg)\frac{1}{M_{\Sigma_{1}}}\bigg(\frac{I_{z}}{ \kappa_{2}(z)^{2}}\bigg)\frac{1}{ (z)^2 \xi(z)},
\end{equation}
 \begin{align}
     W''_{\Sigma_{1}} = \frac{1}{4 \xi(z)}K_{\Sigma_{1}}( z)^{3}\kappa_{1}\left(z\right),
 \end{align}
\begin{align}
    W''_{\Delta L}=\frac{\Gamma_{scattering}}{\textbf{H} z^{2}\xi(z)}.
\end{align}

The solution of Eq. (\ref{Boltz3}) are shown in Fig. (\ref{fig:STG1}) and Fig. (\ref{fig:STG2}). There are two important parameters, $\varphi_{0}$ and $\varphi^{'}_{0}$ that determine the evolution of the scalar field $\varphi$ and therefore the expansion rate of the Universe. In the right panel plot of Fig. (\ref{fig:STG1}) it can be seen that with the decrease in $\varphi_{0}$ the asymmetry production increases. A small initial value of $\varphi_{0}$ leads to a large increase in the expansion rate, and therefore $\Sigma_{1}$ deviates more from its equilibrium abundance, giving rise to an increase in asymmetry. Although the change in $\Sigma_{1}$ abundance is not visible in the left panel plot of Fig. (\ref{fig:STG1}), it is sufficient to increase the asymmetry production. Similarly, from the right panel plot of Fig. (\ref{fig:STG2}) it is observed that the asymmetry production increases with the increase in negative initial value of $\phi^{'}$. Finally, we show a parameter space that generates the observed baryon asymmetry in Fig. (\ref{fig:STG_scan}). We show that is possible to generate the observed baryon asymmetry of the Universe with triplet mass $M_{\Sigma_{1}}\sim 200$ TeV by appropriately choosing the initial value of ($\varphi,\varphi^{'}$).

\section{Conclusion}
\label{section5}

The triplet leptogenesis scenario is motivating as it can have signatures at collider if not too heavy. However, if they are not very heavy they remain in the thermal equilibrum upto a low temperature and sufficient $\rm B-L$ asymmetry can not be produced from their decay. With the standard cosmological history, a sufficient asymmetry can only be generated if the mass of the triplet fermion is very heavy ($M_{\Sigma}\gtrsim 10^{10}$ GeV). Motivating from this we analyze triplet fermion leptogenesis in the context of minimal type-III seesaw model in two non-standard cosmological scenarios. First we consider a scenario with a scalar field, energy density of which falls faster than radiation in the early Universe. Due to the faster expansion, the triplet fermions go out-of equilibrium earlier leading to a sufficient deviation of its abundance from the equilibrium abundance. This significantly increases the asymmetry production. It results in a requirement of relatively lighter $\Sigma_{1}$ to generate the observed baryon asymmetry. We show that by choosing the cosmological parameter appropriately, the scale of leptogenesis can be as low as $5 \rm TeV$. 

In the second scenario, we consider a scalar-tensor theory of gravity where gravity is not described by metric but also by a scalar field. With the appropriate boundary conditions, the scalar field can create a faster expansion of the Universe when it rolls to its minimum. In such a case, the effects of the modified gravity theory are similar to that of a FEU. We show that successful leptogenesis is possible with the lightest triplet mass $M_{\Sigma_1}\sim 200 \rm TeV$. 

The possibility of realizing leptogenesis at TeV or sub TeV scale opens up exciting prospects for experimental verification. In the minimal type-III seesaw model the fermions carry electro-weak gauge charge allowing them to be detected in the collider experiments such as the Large Hadron collider (LHC). These triplet fermions can manifest through their distinctive signatures such as the multi lepton final states accompanied by missing energy, arising from their decay into Standard Model leptons and gauge bosons. For a sub TeV triplet fermion, the current searches at the ATLAS and CMS experiments already put strong constraints on the parameter space. Future high luminosity run at the LHC and next generation colliders could significantly enhanced the constraints to heavier mass.

\section*{Acknowledgment}

D.M. would like to thank the organizers of ``$17^{th}$ International Conference on Interconnections between Particle Physics and Cosmology (PPC) 2024''    at IIT Hyderabad, where the part of this work was discussed. D.M. would also like to thank Pragjyotish College for the travel grant to attend PPC 2024. 

\newpage
\appendix

\section{Scattering cross sections}


\label{Appen:A}

Here, we give the details of the dominant scattering washout processes. The two most significant washout processes are the $\Delta L=2$ violating processes $L H\longrightarrow \bar{L}H^{*}$ (via a $s$-channel $\Sigma_{1}$) and $LL \longrightarrow H^{*}H^{*}$ (via a $t$-channel $\Sigma_{1}$)

\begin{align}
  \hat{\sigma}_{s}(LH \longrightarrow \bar{L}H^{*}) &= \frac{(Y_{\Sigma_{1}}^{\dagger}Y_{\Sigma_{1}})_{11}^2}{4 \pi}\Bigg[2 + x D_{s}^{2 sub} + \bigg(2 - 3x \frac{m_{3}}{\tilde{m_{1}}}\bigg)+ Re\{D_{s}\} + 3\frac{m_{3}}{\tilde{m_{1}}}\bigg(x\frac{m_{3}}{\tilde{m_{1}}}-2\bigg)\\\nonumber
    &-\frac{2}{x}\text{ln}(1+x)\bigg\{1+\bigg(Re\{D_{s}\}+3\frac{m_{3}}{\tilde{m_{1}}} \bigg)(1+x)\bigg\}\Bigg],
\end{align}
where $D_{s}$ is the s channel propagator written as 
\begin{equation}
    D_{s} = \frac{1}{s - M_{\Sigma_{1}}^2 + i \Gamma_{\Sigma_{1}} M_{\Sigma_{1}}}
\end{equation}
and $D_{s}^{2sub}$ is the modulus square of ‘resonance contribution subtracted’ s channel propagator,
expressed in terms of $D_{s}$ as
\begin{equation}
 D_{s}^{2sub} = 1-\frac{\pi}{\Gamma_{\Sigma_{1}} M_{\Sigma_{1}}|D_{s}^2|}\delta(s - M_{\Sigma_{1}}^2).  
\end{equation}
Cross section for t channel process is given by
\begin{equation}
\hat{\sigma}_{t}(LL \longrightarrow H^{*}H^{*}) = \frac{(Y_{\Sigma_{1}}^{\dagger}Y_{\Sigma_{1}})_{11}^2}{4 \pi}\Bigg[\frac{3x}{2}\bigg(\frac{m_{3}^2}{\tilde{m_{1}}^2}+\frac{2}{1+x}\bigg) + \bigg(3\frac{m_{3}}{\tilde{m_{1}}}-\frac{3}{2+x}\bigg)\text{ln}(1+x) \Bigg].    
\end{equation}

\noindent Here $\tilde{m_{1}}$ is the effective neutrino mass parameter defined by

\begin{eqnarray}
    \tilde{m_{1}} = \dfrac{\left(Y_{\Sigma}^{\dagger}Y_{\Sigma}  \right)_{11}v^{2}}{M_{\Sigma_{1}}}.
\end{eqnarray}

\noindent The reaction densities can be determined as follows 

\begin{eqnarray}
    \gamma_{L H \longleftrightarrow \bar{L } H^{*}} & = & \dfrac{T}{64\pi^{4}} \int ds \sqrt{s} \hat{\sigma}_{s}(LH\longrightarrow \bar{L}H^{*})\kappa_{1}(\sqrt{s}/T) \\ 
     \gamma_{L L \longleftrightarrow H^{*} H^{*}} & = & \dfrac{T}{64\pi^{4}} \int ds \sqrt{s} \hat{\sigma}_{t}(LL\longrightarrow H^{*} H^{*})\kappa_{1}(\sqrt{s}/T) .
\end{eqnarray}

\noindent The rate of scattering washout is defined as

\begin{eqnarray}
    \Gamma_{scatterings}=\dfrac{\gamma_{l H \longrightarrow \bar{L}H^{*}}+\gamma_{LL \longrightarrow H^{*}H^{*}}}{n_{l}^{eq}},
\end{eqnarray}

\noindent $n_{l}^{eq}$ is the equilibrium number density of the leptons.




-








\newpage

\bibliographystyle{JHEP}
\bibliography{ref.bib}

\end{document}